\documentclass[preprint,authoryear]{elsarticle}

\usepackage{url}
\usepackage{booktabs}
\usepackage{times}
\usepackage{amsthm}
\usepackage{amssymb}
\usepackage{amsmath}
\usepackage{array}
\usepackage[all]{xy}
\usepackage{subfig}
\usepackage{color}
\usepackage{colortbl}
\usepackage[noend]{algorithmic}
\usepackage{algorithm}
\usepackage{multicol}
\usepackage{caption}

\definecolor{grey}{rgb}{0.9,0.9,0.9}

\newcommand{\func}[1]{\ensuremath{\mathit{#1}}}
\newcommand{\var}[1]{\ensuremath{\mathit{#1}}}

\newcommand{\ImprovePath}{\func{ImprovePath}}
\newcommand{\Cluster}{\func{Cluster}}

\newcommand{\instance}[1]{\texttt{#1}}

\newcommand{\heuristic}[1]{\textsf{#1}}

\newcommand{\CO}{\heuristic{CO}}
\newcommand{\MA}{\heuristic{ma}}
\newcommand{\NN}{\heuristic{NN}}
\newcommand{\LK}{\heuristic{LK}}
\newcommand{\LKtsp}{\heuristic{LK$_\text{tsp}$}}

\newcommand{\opt}[3]{\heuristic{#1-opt}$_\text{#2}^\text{#3}$}
\newcommand{\optshort}[3]{\heuristic{#1o}$_\text{#2}^\text{#3}$}

\newcommand{\LKb}[3]{\heuristic{B}\ensuremath{_\text{#1}^\text{#2#3}}}
\newcommand{\LKc}[3]{\heuristic{C}\ensuremath{_\text{#1}^\text{#2#3}}}
\newcommand{\LKs}[3]{\heuristic{S}\ensuremath{_\text{#1}^\text{#2#3}}}
\newcommand{\LKe}[2]{\heuristic{E}\ensuremath{_\text{#1}^\text{#2}}}

\newtheorem{theorem}{Theorem}

\begin{document}

\title{Lin-Kernighan Heuristic Adaptations for the Generalized Traveling Salesman Problem}

\author[rhul]{D.~Karapetyan\corref{cor1}}
\ead{daniel.karapetyan@gmail.com}
\author[rhul]{G.~Gutin}
\ead{gutin@cs.rhul.ac.uk}

\cortext[cor1]{Corresponding author}
\address[rhul]{Royal Holloway London University, Egham, Surrey, TW20 0EX, United Kingdom}


\date{}

\begin{abstract}
The Lin-Kernighan heuristic is known to be one of the most successful heuristics for the Traveling Salesman Problem (TSP)\@.  It has also proven its efficiency in application to some other problems.

In this paper we discuss possible adaptations of TSP heuristics for the Generalized Traveling Salesman Problem (GTSP) and focus on the case of the Lin-Kernighan algorithm.  At first, we provide an easy-to-understand description of the original Lin-Kernighan heuristic.  Then we propose several adaptations, both trivial and complicated.  Finally, we conduct a fair competition between all the variations of the Lin-Kernighan adaptation and some other GTSP heuristics.

It appears that our adaptation of the Lin-Kernighan algorithm for the GTSP reproduces the success of the original heuristic.  Different variations of our adaptation outperform all other heuristics in a wide range of trade-offs between solution quality and running time, making Lin-Kernighan the state-of-the-art GTSP local search.

\end{abstract}

\maketitle

\section{Introduction}

One of the most successful heuristic algorithms for the famous Traveling Salesman Problem (TSP) known so far is the Lin-Kernighan heuristic~\citep{Lin1973}.  It was proposed almost forty years ago but even nowadays it is the state-of-the-art TSP local search~\citep{Johnson2002}.

In this paper we attempt to reproduce the success of the original TSP Lin-Kernighan heuristic for the Generalized Traveling Salesman Problem (GTSP), which is an important extension of TSP\@.  In the TSP, we are given a set $V$ of $n$ vertices and weights $w(x \to y)$ of moving from a vertex $x \in V$ to a vertex $y \in V$.  A feasible solution, or a tour, is a cycle visiting every vertex in $V$ exactly once.  In the GTSP, we are given a set $V$ of $n$ vertices, weights $w(x \to y)$ of moving from $x \in V$ to $y \in V$ and a partition of $V$ into $m$ nonempty clusters $C_1, C_2, \ldots, C_m$ such that $C_i \cap C_j = \varnothing$ for each $i \neq j$ and $\bigcup_i C_i = V$\@.  A feasible solution, or a tour, is a cycle visiting exactly one vertex in every cluster.  The objective of both TSP and GTSP is to find the shortest tour.

If the weight matrix is symmetric, i.e., $w(x \to y) = w(y \to x)$ for any $x, y \in V$, the problem is called \emph{symmetric}.  Otherwise it is an \emph{asymmetric} GTSP\@.  In what follows, the number of vertices in cluster $C_i$ is denoted as $|C_i|$, the size of the largest cluster is $s$, and $\Cluster(x)$ is the cluster containing a vertex $x$.  The weight function $w$ can be used for edges, paths $w(x_1 \to x_2 \to \ldots \to x_k) = w(x_1 \to x_2) + w(x_2 \to x_3) + \ldots + w(x_{k-1} \to x_k)$, and cycles.

Since Lin-Kernighan is designed for the symmetric problem, we do not consider the asymmetric GTSP in this research.  However, some of the algorithms proposed in this paper are naturally suited for both symmetric and asymmetric cases.

Observe that the TSP is a special case of the GTSP when $|C_i| = 1$ for each $i$ and, hence, the GTSP is NP-hard.  The GTSP has a host of applications in warehouse order picking with multiple stock locations, sequencing computer files, postal routing, airport selection and routing for courier planes and some others, see, e.g., \citep{Fischetti1995,Fischetti1997,Laporte1996,Noon1991} and references therein.

A lot of attention was paid in the literature to solving the GTSP\@.  Several researchers \citep{Ben-Arieh2003,Laporte1999,Noon1993} proposed transformations of the GTSP into the TSP\@.  At first glance, the idea to transform a little-studied problem into a well-known one seems to be natural; however, this approach has a very limited application.  On the one hand, it requires exact solutions of the obtained TSP instances because even a near-optimal solution of such TSP may correspond to an infeasible GTSP solution.  On the other hand, the produced TSP instances have quite an unusual structure which is difficult for the existing solvers.  A more efficient way to solve the GTSP exactly is a branch-and-bound algorithm designed by~\citet{Fischetti1997}\@.  This algorithm was able to solve instances with up to 89 clusters.  Two approximation algorithms were proposed in the literature, but both of them are unsuitable for the general case of the problem, and the guarantied solution quality is unreasonably low for real-world applications, see~\citep{Bontoux2009} and references therein.

In order to obtain good (i.e., not necessarily exact) solutions for larger GTSP instances, one should use the heuristic approach.  Several construction heuristics and local searches were discussed in~\citep{Bontoux2009,GK_GTSP_GA_2008,Hu2008,Renaud1998,Snyder2000} and some others.  A number of metaheuristics were proposed by \citet{Bontoux2009,GK_GTSP_GA_2008,GK_GTSP_GA_2007,Huang2005,Pintea2007,Silberholz2007,Snyder2000,Tasgetiren2007,Yang2008}.

In this paper we thoroughly discuss possible adaptations of a TSP heuristic for the GTSP and focus on the Lin-Kernighan algorithm.  The idea of the Lin-Kernighan algorithm was already successfully applied to the Multidimensional Assignment Problem~\citep{Balas1991,GK_MAP_LS_2010}.  A straightforward adaptation for the GTSP was proposed by~\citet{Hu2008}; their algorithm constructs a set of TSP instances and solves all of them with the TSP Lin-Kernighan heuristic.  \citet{Bontoux2009} apply the TSP Lin-Kernighan heuristic to the TSP tours induced by the GTSP tours.  It will be shown in Section~\ref{sec:gtsp_adaptation} that both of these approaches are relatively weak.

The Lin-Kernighan heuristic is a sophisticated algorithm adjusted specifically for the TSP\@.  The explanation provided by~\citet{Lin1973} is full of details which complicate understanding of the main idea of the method.  We start our paper from a clear explanation of a simplified TSP Lin-Kernighan heuristic (Section~\ref{sec:lin_kernighan}) and then propose several adaptations of the heuristic for the GTSP (Section~\ref{sec:gtsp_adaptation}).  In Section~\ref{sec:experiments}, we provide results of a thorough experimental evaluation of all the proposed Lin-Kernighan adaptations and discuss the success of our approach in comparison to other GTSP heuristics.  In Section~\ref{sec:conclusion} we discuss the outcomes of the conducted research and select the state-of-the-art GTSP local searches.

\section{The TSP Lin-Kernighan Heuristic}
\label{sec:lin_kernighan}

In this section we describe the TSP Lin-Kernighan heuristic (\LKtsp)\@.  It is a simplified version of the original algorithm.  Note that~\citep{Lin1973} was published almost 40 years ago, when modest computer resources, obviously, influenced the algorithm design, hiding the main idea behind the technical details.  Also note that, back then, the `goto' operator was widely used; this affects the original algorithm description.  In contrast, our interpretation of the algorithm is easy to understand and implement.

\LKtsp{} is a generalization of the $k$-opt local search.  The $k$-opt neighborhood $N_\text{$k$-opt}(T)$ includes all the TSP tours which can be obtained by removing $k$ edges from the original tour $T$ and adding $k$ different edges such that the resulting tour is feasible.  Observe that exploring the whole $N_\text{$k$-opt}(T)$ takes $O(n^k)$ operations and, thus, with a few exceptions, only 2-opt and rarely 3-opt are used in practice~\citep{Johnson2002,Rego2006}.

Similarly to $k$-opt, \LKtsp{} tries to remove and insert edges in the tour but it explores only some parts of the $k$-opt neighborhood that deem to be the most promising.  Consider removing an edge from a tour; this produces a path.  Rearrange this path to minimize its weight.  To close up the tour we only need to add one edge.  Since we did not consider this edge during the path optimization, it is likely that its weight is neither minimized nor maximized.  Hence, the weight of the whole tour is probably reduced together with the weight of the path.  Here is a general scheme of \LKtsp{}:
\begin{enumerate}
	\item Let $T$ be the original tour.
	\item \label{item:nextedge} For every edge $e \to b \in T$ do the following:
	\begin{enumerate}
		\item Let $P = b \to \ldots \to e$ be the path obtained from $T$ by removing the edge $e \to b$.
		\item Rearrange $P$ to minimize its weight.  Every time an improvement is found during this optimization, try to close up the path $P$\@.  If it leads to a tour shorter than $T$, save this tour as $T$ and start the whole procedure again.
		\item If no tour improvement was found, continue to the next edge (Step~\ref{item:nextedge}).
	\end{enumerate}
\end{enumerate}

\begin{figure}[ht]
\centering  

\subfloat[The original path.]{
\label{fig:original_path}
\xymatrix@R=3em@C=3em@L=0.1em{
	&	*++[o][F-]{b} \ar@{->}[r]
	&	*++[o][F-]{\phantom{1}} \ar@{->}[r]
	&	*++[o][F-]{x} \ar@{->}[r]^{w(x \to y)}
	&	*++[o][F-]{y} \ar@{->}[r]
	&	*++[o][F-]{\phantom{1}} \ar@{->}[r]
	&	*++[o][F-]{e}
}}
\\[1em]
\subfloat[The path after a local search move.]{
\label{fig:rejoined_path}
\xymatrix@R=3em@C=3em@L=0.1em{
	&	*++[o][F-]{b} \ar@{->}[r]
	&	*++[o][F-]{\phantom{1}} \ar@{->}[r]
	&	*++[o][F-]{x}	\ar@{--}[r]
	&	*++[o][F-]{y} \ar@{<-}[r]
	&	*++[o][F-]{\phantom{1}} \ar@{<-}[r]
	&	*++[o][F-]{e} \ar@/_3em/@{<-}[lll]_{w(x \to e)}
}}

\caption{An example of a local search move for a path improvement.  The weight of the path is reduced by $w(x \to y) - w(x \to e)$.}

\label{fig:path_local_search}
\end{figure}
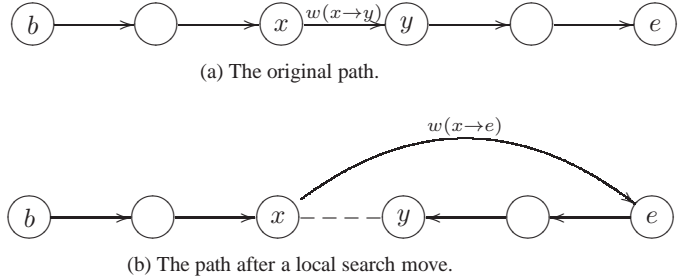

In order to reduce the weight of the path, a local search is used as follows.  On every move, it tries to break up the path into two parts, invert one of these parts, and then rejoin them (see Figure~\ref{fig:path_local_search}).
In particular, the algorithm tries every edge $x \to y$ and selects the one which maximizes the gain $g = w(x \to y) - w(e \to x)$.  If the maximum $g$ is positive, the corresponding move is an improvement and the local search is applied again to the improved path.

Observe that this algorithm tries only the best improvement and skips the other ones.  A natural enhancement of the heuristic would be to use a backtracking mechanism to try all the improvements.  However, this would slow down the algorithm too much.  A compromise is to use the backtracking only for the first $\alpha$ moves.  This approach is implemented in a recursive function $\ImprovePath(P, \var{depth}, R)$, see Algorithm~\ref{alg:improvepath_scheme}.

\begin{algorithm}[ht]
\caption{$\ImprovePath{}(P, \var{depth}, R)$ recursive algorithm (\LKtsp{} version).  The function either terminates after an improved tour is found or finishes normally with no profit.}
\label{alg:improvepath_scheme}
\begin{algorithmic}
\REQUIRE The path $P = b \to \ldots \to e$, recursion depth \var{depth} and a set of restricted vertices $R$.
\IF {$\var{depth} < \alpha$}
	\FOR {every edge $x \to y \in P$ such that $x \notin R$}
		\STATE Calculate $g = w(x \to y) - w(e \to x)$ (see Figure~\ref{fig:rejoined_path}).
		\IF {$g > 0$}
			\IF {the tour $b \to \ldots \to x \to e \to \ldots \to y \to b$ is an improvement over the original one}
				\STATE Accept the produced tour and \textbf{terminate}.
			\ELSE
				\STATE $\ImprovePath(b \to \ldots \to x \to e \to \ldots \to y, \var{depth} + 1, R \cup \{ x \})$.
			\ENDIF
		\ENDIF
	\ENDFOR
\ELSE
	\STATE Find the edge $x \to y$ which maximizes $g = w(x \to y) - w(e \to x)$.
	\IF {$g > 0$}
		\IF {the tour $b \to \ldots \to x \to e \to \ldots \to y \to b$ is an improvement over the original one}
			\STATE Accept the produced tour and \textbf{terminate}.
		\ELSE
			\RETURN $\ImprovePath{}(b \to \ldots \to x \to e \to \ldots \to y, \var{depth} + 1, R \cup \{ x \})$.
		\ENDIF
	\ENDIF
\ENDIF
\end{algorithmic}
\end{algorithm}

$\ImprovePath(P, 1, \varnothing)$ takes $O(n^\alpha \cdot \var{depth}_{\max})$ operations, where $\var{depth}_{\max}$ is the maximum depth of recursion achieved during the run.  Hence, one should use only small values of backtracking depth $\alpha$.

\bigskip

The algorithm presented above is a simplified Lin-Kernighan heuristic.  Here is a list of major differences between the described algorithm and the original one.
\begin{enumerate}
	\item The original heuristic does not accept the first found tour improvement.  It records it and continues optimizing the path in the hope of finding a better tour improvement.  Note that it was reported by~\citet{Helsgaun2000} that this complicates the algorithm but does not really improve its quality.
	
	\item The original heuristic does not try all the $n$ options when optimizing a path.  It considers only the five shortest edges $x \to e$ in the non-decreasing order.  This hugely reduces the running time and helps to find the best rather than the first improvement on the backtracking stage.  However, this speed-up approach is known to be a weak point of the original implementation~\citep{Helsgaun2000,Johnson2002}.  Indeed, even if the edge $x \to y$ is long, the algorithm does not try to break it if the edge $x \to e$ is not in the list of five shortest edges to $e$.	 
	
	Note that looking for the closest vertices or clusters may be meaningless in the application to the GTSP\@.  In our implementation, every edge $x \to y$ is considered.

	\item The original heuristic does not allow deleting the previously added edges or adding the previously deleted edges.  It was noted~\citep{Helsgaun2000,Johnson2002} that either of these restrictions is enough to prevent an infinite loop.  In our implementation a previously deleted edge is allowed to be added again but every edge can be deleted only once.  Our implementation also prevents some other moves; however, the experimental evaluation shows that this does not affect the performance of the heuristic.
	
	\item The original heuristic also considers some more sophisticated moves to produce a path from the tour.
	
	\item The original heuristic is, in fact, embedded into a metaheuristic which runs the optimization several times.  There are several tricks related to the metaheuristic which are inapplicable to a single run.
\end{enumerate}

The worst case time complexity of the Lin-Kernighan heuristic seems to be unknown from the literature \citep{Helsgaun2009} but we assume that it is exponential.  Indeed, observe that the number of iterations of the $k$-opt local search may be non-polynomial for any $k$~\citep{Chandra1994} and that \LKtsp{} is a modification of $k$-opt.  However, \citet{Helsgaun2009} notes that such undesirable instances are very rare and normally \LKtsp{} proceeds in a polynomial time.

\section{Adaptations of the Lin-Kernighan Heuristic for the GTSP}
\label{sec:gtsp_adaptation}

It may seem that the GTSP is only a slight variation of the TSP\@.  In particular, one may propose splitting the GTSP into two problems~\citep{Renaud1998}: solving the TSP induced by the given tour to find the cluster order, and finding the shortest cycle visiting the clusters according to the found order.  We will show now that this approach is poor with regards to solution quality.  Let $N_\text{TSP}(T)$ be a set of tours which can be obtained from the tour $T$ by reordering the vertices in $T$.  Observe that one has to solve a TSP instance induced by $T$ to find the best tour in $N_\text{TSP}(T)$.  

Let $N_\text{CO}(T)$ be a set of all the GTSP tours which visit the clusters in exactly the same order as in $T$.  The size of the $N_\text{CO}(T)$ neighborhood is $\prod_{i=1}^m |C_i| \in O(s^m)$ but there exists a polynomial algorithm (we call it \emph{Cluster Optimization}, \CO) which finds the best tour in $N_\text{CO}(T)$ in $O(m s^3)$ operations \citep{Fischetti1997}.  Moreover, it requires only $O(m s^2 \cdot \min_i |C_i|)$ time, i.e., if the instance has at least one cluster of size $O(1)$, \CO{} proceeds in $O(m s^2)$.  (Recall that $s$ is the size of the largest cluster: $s = \max_i |C_i|$.)

The following theorem shows that splitting the GTSP into two problems (local search in $N_\text{TSP}(T)$ and then local search in $N_\text{CO}(T)$) does not guarantee any solution quality.

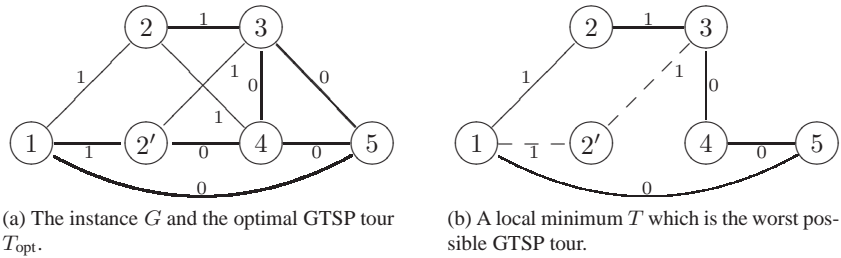
\begin{figure}[ht]
\centering  
\subfloat[The instance $G$ and the optimal GTSP tour $T_\text{opt}$.]{
\label{fig:gtsp_problem}
\xymatrix@R=2.5em@C=2.5em@L=0.1em{
	&	*++[o][F-]{2}		\ar@{-}[r]^1 
										\ar@{-}[rd]_(0.7)1
	&	*++[o][F-]{3}		\ar@{-}[rd]^0 
										\ar@{-}@<0.5\linethickness>[rd] 
										\ar@{-}@<-0.5\linethickness>[rd]
\\
		*++[o][F-]{1}		\ar@{-}[ru]^1 
										\ar@{-}[r]_1
										\ar@{-}@<0.5\linethickness>[r]
										\ar@{-}@<-0.5\linethickness>[r]
	&	*++[o][F-]{2'}	\ar@{-}[ru]_(0.7)1 
										\ar@{-}[r]_0
										\ar@{-}@<0.5\linethickness>[r]
										\ar@{-}@<-0.5\linethickness>[r]
	&	*++[o][F-]{4}		\ar@{-}[r]_0 
										\ar@{-}[u]^0
										\ar@{-}@<0.5\linethickness>[u]
										\ar@{-}@<-0.5\linethickness>[u]
	&	*++[o][F-]{5}		\ar@/^2em/@{-}[lll]_0
										\ar@/^2em/@{-}@<0.5\linethickness>[lll]
										\ar@/^2em/@{-}@<-0.5\linethickness>[lll]
}}
\qquad
\subfloat[A local minimum $T$ which is the worst possible GTSP tour.]{
\label{fig:worst_tour}
\xymatrix@R=2.5em@C=2.5em@L=0.1em{
	&	*++[o][F-]{2}		\ar@{-}[r]^1
	&	*++[o][F-]{3}		\ar@{-}[d]^0
\\
		*++[o][F-]{1}		\ar@{-}[ru]^1 
										\ar@{--}[r]_1
	&	*++[o][F-]{2'}	\ar@{--}[ru]_(0.7)1
	&	*++[o][F-]{4}		\ar@{-}[r]_0
	&	*++[o][F-]{5}		\ar@/^2em/@{-}[lll]_0
}}

\caption{An example of a local minimum in both $N_\text{TSP}(T)$ and $N_\text{CO}(T)$ which is a longest possible GTSP tour.}

\label{fig:tsp_co_local_minimum}
\end{figure}

\begin{theorem}
\label{th:tsp_co_local_minimum}
The best tour among $N_\text{CO}(T) \cup N_\text{TSP}(T)$ can be a longest GTSP tour different from a shortest one.
\end{theorem}
\proof
Consider the GTSP instance $G$ in Figure~\ref{fig:gtsp_problem}.  It is a symmetric GTSP containing 5 clusters $\{ 1 \}$, $\{ 2, 2' \}$, $\{ 3 \}$, $\{ 4 \}$ and $\{ 5 \}$.  The weights of the edges not displayed in the graph are as follows: $w(1 \to 3) = w(1 \to 4) = 0$ and $w(2 \to 5) = w(2' \to 5) = 1$.

Observe that the tour $T = 1 \to 2 \to 3 \to 4 \to 5 \to 1$, shown in Figure~\ref{fig:worst_tour}, is a local minimum in both $N_\text{CO}(T)$ and $N_\text{TSP}(T)$.  The dashed line shows the second solution in $N_\text{CO}(T)$ but it gives the same objective value.  It is also clear that $T$ is a local minimum in $N_\text{TSP}(T)$.  Indeed, all the edges incident to the vertex 2 are of weight 1, and, hence, any tour through the vertex 2 is at least of weight 2.

The tour $T$ is in fact a longest tour in $G$\@.  Observe that all nonzero edges in $G$ are incident to the vertices 2 and $2'$.  Since only one of these vertices can be visited by a tour, at most two nonzero edges can be included into a tour.  Hence, the weight of the worst tour in $G$ is 2.

However, there exists a better GTSP tour $T_\text{opt} = 1 \to 2' \to 4 \to 3 \to 5 \to 1$ of weight 1, see Figure~\ref{fig:gtsp_problem}.

\qed

In fact, the TSP and the GTSP behave quite differently during optimization.  Observe that there exists no way to find out quickly if some modification of the cluster order improves the tour.  Indeed, choosing wrong vertices within clusters may lead to an arbitrary large increase of the tour weight.  And since a replacement of a vertex within one cluster may require a replacement of vertices in the neighbor clusters, any local change influences the whole tour in general case.

\subsection{Local Search Adaptation}
\label{sec:ls_adaptation}

A typical local search with the neighborhood $N(T)$ performs as follows:
\begin{algorithmic}[ht]
\REQUIRE The original solution $T$.
\FORALL {$T' \in N(T)$}
	\IF {$w(T') < w(T)$}
		\STATE $T \gets T'$.
		\STATE Run the whole algorithm again.
	\ENDIF
\ENDFOR
\RETURN $T$.
\end{algorithmic}

Let $N_1(T) \subseteq N_\text{TSP}(T)$ be a neighborhood of some TSP local search $\func{LS}_1(T)$.  Let $N_2(T) \subseteq N_\text{CO}(T)$ be a neighborhood of some GTSP local search $\func{LS}_2(T)$ which leaves the cluster order fixed.  Then one can think of the following two adaptations of a TSP local search for the GTSP:
\begin{enumerate}[(i)]
	\item \label{item:coinsidetsp} Enumerate all solutions $T' \in N_1(T)$.  For every candidate $T'$ run $T' \gets \func{LS}_2(T')$ to optimize it in $N_2(T')$.
	\item \label{item:tspinsideco} Enumerate all solutions $T' \in N_2(T)$.  For every candidate $T'$ run $T' \gets \func{LS}_1(T')$ to optimize it in $N_1(T')$.
\end{enumerate}

Observe that the TSP neighborhood $N_1(T)$ is normally harder to explore than the cluster optimization neighborhood $N_2(T)$.  Consider, e.g., $N_1(T) = N_\text{TSP}(T)$ and $N_2(T) = N_\text{CO}(T)$.  Then both options yield an optimal GTSP solution but Option~(\ref{item:coinsidetsp}) requires $O(m! m s^3)$ operations while Option~(\ref{item:tspinsideco}) requires $O(s^m m!)$ operations.

Moreover, many practical applications of the GTSP have some localization of clusters, i.e., $|w(x \to y_1) - w(x \to y_2)| \ll w(x \to y_1)$ on average, where $\Cluster(y_1) = \Cluster(y_2) \neq \Cluster(x)$.  Hence, the landscape of $N_2(T)$ depends on the cluster order more than the landscape of $N_1(T)$ depends on the vertex selection.  From above it follows that Option~(\ref{item:coinsidetsp}) is preferable.

Option~(\ref{item:tspinsideco}) was used by~\citet{Hu2008} as follows.  The cluster optimization neighborhood $N_2(T)$ includes there all the tours which differ from $T$ in exactly one vertex.  For every $T' \in N_2(T)$ the Lin-Kernighan heuristic was applied.  This results in $n$ runs of the Lin-Kernighan heuristic which makes the algorithm unreasonably slow.

Option~(\ref{item:coinsidetsp}) may be implemented as follows:
\begin{algorithmic}[ht]
\REQUIRE The original tour $T$.
\FORALL {$T' \in N_1(T)$}
	\STATE $T' \gets \func{QuickImprove}(T')$.
	\IF {$w(T') < w(T)$}
		\STATE $T \gets \func{SlowImprove}(T')$.
		\STATE Run the whole algorithm again.
	\ENDIF
\ENDFOR
\RETURN $T$.
\end{algorithmic}
Here $\func{QuickImprove}(T)$ and $\func{SlowImprove}(T)$ are some tour improvement heuristics which leave the cluster order unchanged.  Formally, these heuristics should meet the following requirements: 
\begin{itemize}
	\item $\func{QuickImprove}(T), \func{SlowImprove}(T) \in N_\text{CO}(T)$ for any tour $T$;
	\item $w(\func{QuickImprove}(T)) \le w(T)$ and $w(\func{SlowImprove}(T)) \le w(T)$ for any tour $T$.
\end{itemize}
\func{QuickImprove} is applied to every candidate $T'$ before its evaluation.  \func{SlowImprove} is only applied to successful candidates in order to further improve them.  One can think of the following improvement functions: 
\begin{itemize}
	\item Trivial $I(T)$ which leaves the solution without any change: $I(T) = T$.
	\item Full optimization $\mathit{CO}(T)$ which applies the \CO{} algorithm to the given solution.
	\item Local optimization $L(T)$.  It updates the vertices only within clusters, affected by the latest solution change.  E.g., if a tour $x_1 \to x_2 \to x_3 \to x_4 \to x_1$ was changed to $x_1 \to x_3 \to x_2 \to x_4 \to x_1$, some implementation of $L(T)$ will try every $x_1 \to x'_3 \to x'_2 \to x_4 \to x_1$, where $x'_2 \in \Cluster(x_2)$ and $x'_3 \in \Cluster(x_3)$.
\end{itemize}

There are five meaningful combinations of \func{QuickImprove} and \func{SlowImprove}:
\begin{enumerate}
	\item \label{item:plain} $\func{QuickImprove}(T) = I(T)$ and $\func{SlowImprove}(T) = I(T)$.  This actually yields the original TSP local search.

	\item \label{item:applyco} $\func{QuickImprove}(T) = I(T)$ and $\func{SlowImprove}(T) = \func{CO}(T)$, i.e., the algorithm explores the TSP neighborhood but every time an improvement is found, the solution $T$ is optimized in $N_\text{CO}(T)$.  One can also consider $\func{SlowImprove}(T) = L(T)$, but it has no practical interest.  Indeed, \func{SlowImprove} is used quite rarely and so its impact on the total running time is negligible.  At the same time, $\func{CO}(T)$ is much better than $L(T)$ with respect to solution quality.

	\item \label{item:vary} $\func{QuickImprove}(T) = L(T)$ and $\func{SlowImprove}(T) = I(T)$, i.e., every solution $T' \in N(T)$ is improved locally before it is compared to the original solution.

	\item \label{item:varyco} $\func{QuickImprove}(T) = L(T)$ and $\func{SlowImprove}(T) = \func{CO}(T)$, which is the same as Option~\ref{item:vary} but it additionally optimizes the solution $T'$ globally in $N_\text{CO}(T')$ every time an improvement is found.

	\item \label{item:exact} $\func{QuickImprove}(T) = \func{CO}(T)$ and $\func{SlowImprove}(T) = I(T)$, i.e., every candidate $T' \in N(T)$ is optimized globally in $N_\text{CO}(T')$ before it is compared to the original solution $T$.
\end{enumerate}

These adaptations were widely applied in the literature.  For example, the heuristics G2 and G3~\citep{Renaud1998} are actually 2-opt and 3-opt adapted according to Option~\ref{item:exact}.  An improvement over the naive implementation of 2-opt adapted in this way is proposed by~\citet{Hu2008}; asymptotically, it is faster by factor 3.  However, this approach is still too slow.  Adaptations of 2-opt and some other heuristics according to Option~\ref{item:vary} were used by \citet{Fischetti1997}, \citet{GK_GTSP_GA_2008}, \citet{Silberholz2007}, \citet{Snyder2000}, and \citet{Tasgetiren2007}.  Some unadapted TSP local searches (Option~\ref{item:plain}) were used by \citet{Bontoux2009}, \citet{GK_GTSP_GA_2008}, \citet{Silberholz2007}, and \citet{Snyder2000}.

\subsection{Adaptation of \LKtsp{}}
\label{sec:lk_adaptation}

In this section we present our adaptation \LK{} of \LKtsp{} for the GTSP\@.  A pseudo-code of the whole heuristic is presented in Algorithm~\ref{alg:lk_main}.
\begin{algorithm}[!ht]
\caption{LK general implementation}
\label{alg:lk_main}

\begin{algorithmic}
\REQUIRE The original tour $T$.
\STATE Initialize the number of idle iterations $i \gets 0$.
\WHILE {$i < m$}
	\STATE Cyclically select the next edge $e \to b \in T$.
	\STATE Let $P_o = b \to \ldots \to e$ be the path obtained from $T$ by removing the edge $e \to b$.
	\STATE Run $T' \gets \func{ImprovePath}(P_o, 1, \varnothing)$ (see below).
	\IF {$w(T') < w(T)$}
		\STATE Set $T = \func{ImproveTour}(T')$.
		\STATE Reset the number of idle iterations $i \gets 0$.
	\ELSE
		\STATE Increase the number of idle iterations $i \gets i + 1$.
	\ENDIF
\ENDWHILE
\end{algorithmic}

\rule{\linewidth}{\heavyrulewidth}
\textbf{Procedure} $\func{ImprovePath}(P, \var{depth}, R)$\\
\rule[1em]{\linewidth}{\linethickness}

\vspace{-3ex}
\begin{algorithmic}
\REQUIRE The path $P = b \to \ldots \to e$, recursion depth $\var{depth}$ and the set of restricted vertices $R$.
\IF {$\var{depth} \ge \alpha$}
	\STATE Find the edge $x \to y \in P$, $x \neq b$, $x \notin R$ such that it maximizes the path gain $\mathit{Gain}(P, x \to y)$.
\ELSE
	\STATE Repeat the rest of the procedure for every edge $x \to y \in P$, $x \neq b$, $x \notin R$.
\ENDIF \bigskip

\STATE Conduct the local search move: $P \gets \func{RearrangePath}(P,\ x \to y)$.
\IF {$\func{GainIsAcceptable}(P,\ x \to y)$}
	\STATE Replace the edge $x \to y$ with $x \to e$ in $P$.
	
	\STATE $T' = \func{CloseUp}(P)$.
	\IF {$w(T') \ge w(T)$}
		\STATE Run $T' \gets \func{ImprovePath}(P, \var{depth} + 1, R \cup \{ x \})$.
	\ENDIF
	
	\IF {$w(T') < w(T)$}
		\RETURN $T'$.
	\ELSE
		\STATE Restore the path $P$.
	\ENDIF
\ENDIF
\RETURN $T$.
\end{algorithmic}
\end{algorithm}
Some of its details are encapsulated into the following functions (note that \LKtsp{} is not a typical local search based on some neighborhood and, thus, the framework presented above cannot be applied to it straightforwardly):
\begin{itemize}
	\item $\func{Gain}(P,\ x \to y)$ is intended to calculate the gain of breaking a path $P$ at an edge $x \to y$.
	
	\item $\func{RearrangePath}(P,\ x \to y)$ removes an edge $x \to y$ from a path $P$ and adds the edge $x \to e$, where $P = b \to \ldots \to x \to y \to \ldots \to e$, see Figure~\ref{fig:path_local_search}.  Together with \func{CloseUp}, it includes an implementation of $\func{QuickImprove}(T)$ (see Section~\ref{sec:ls_adaptation}), so \func{RearrangePath} may also apply some cluster optimization.
	
	\item $\func{GainIsAcceptable}(P,\ x \to y)$ determines if the gain of breaking a path $P$ at an edge $x \to y$ is worth any further effort.
	
	\item $\func{CloseUp}(P)$ adds an edge to a path $P$ to produce a feasible tour.  Together with \func{RearrangePath}, it includes an implementation of $\func{QuickImprove}(T)$ (see Section~\ref{sec:ls_adaptation}), so \func{CloseUp} may also apply some cluster optimization.
	
	\item $\func{ImproveTour}(T)$ is a tour improvement function.  It is an analogue to $\func{SlowImprove}(T)$ (see Section~\ref{sec:ls_adaptation}).
\end{itemize}

These functions are the key points in the adaptation of \LKtsp{} for the GTSP\@.  They determine the behaviour of the heuristic.  In Sections~\ref{sec:basic}, \ref{sec:closest_shortest} and~\ref{sec:exact} we describe different implementations of these functions.

\subsection{The Basic Variation}
\label{sec:basic}

The \heuristic{Basic} variation of \LKtsp{} (in what follows denoted by \LKb{}{}{}) is a trivial adaptation of LK according to Option~\ref{item:plain} (see Section~\ref{sec:ls_adaptation}).  It defines the functions \func{Gain}, \func{RearrangePath}, \func{CloseUp} and \func{ImproveTour} as follows:
$$
\func{Gain}_\text{B}(b \to \ldots \to e,\ x \to y) = w(x \to y) - w(e \to x) \,,
$$
$$
\func{RearrangePath}_\text{B}(b \to \ldots \to x \to y \to \ldots \to e,\ x \to y) = b \to \ldots \to x \to e \to \ldots \to y \,,
$$
$$
\func{CloseUp}_\text{B}(b \to \ldots \to e) = b \to \ldots \to e \to b \,,
$$
and $\func{ImproveTour}_\text{B}(T)$ is trivial.  We also consider a \LKb{}{}{co} variation (Option~\ref{item:applyco}) which applies \CO{} every time an improvement is found: $\func{ImproveTour}(T) = \func{CO}(T)$.

The implementation of $\func{GainIsAcceptable}(G, P)$ will be discussed in Section~\ref{sec:gain}.

\subsection{The Closest and the Shortest Variations}
\label{sec:closest_shortest}

The \heuristic{Closest} and the \heuristic{Shortest} variations (denoted as \LKc{}{}{} and \LKs{}{}{}, respectively) are two adaptations of \LKtsp{} according to Option~\ref{item:vary}, i.e., $\func{QuickImprove}(T) = L(T)$ and $\func{SlowImprove}(T) = I(T)$.  In other words, a local cluster optimization is applied to every candidate during the path optimization.

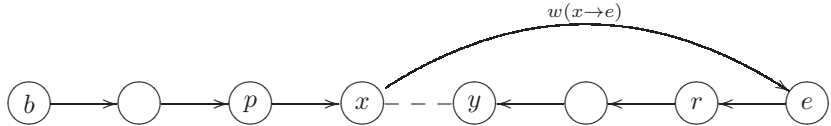
\begin{figure}[ht]
\centerline{
\noindent\xymatrix@R=2.5em@C=2.5em@L=0.1em{
		*++[o][F-]{b} \ar@{->}[r]
	&	*++[o][F-]{\phantom{1}} \ar@{->}[r]
	&	*++[o][F-]{p} \ar@{->}[r]
	&	*++[o][F-]{x}	\ar@{--}[r]
	&	*++[o][F-]{y} \ar@{<-}[r]
	&	*++[o][F-]{\phantom{1}} \ar@{<-}[r]
	&	*++[o][F-]{r} \ar@{<-}[r]
	&	*++[o][F-]{e} \ar@/_3em/@{<-}[llll]_{w(x \to e)}
}}
\caption{Path optimization.}
\label{fig:path_vary}
\end{figure}

Consider an iteration of the path improvement heuristic \func{ImprovePath}.  Let the path $P = b \to \ldots \to x \to y \to \ldots \to e$ be broken at the edge $x \to y$ (see Figure~\ref{fig:path_vary}).
Then, to calculate $\func{Gain}(P,\ x \to y)$ in \LKc{}{}{}, we replace $x \in X$ with $x' \in X$ such that the edge $x \to e$ is minimized:
\begin{multline*}
\func{Gain}_\text{C}(b \to \ldots \to p \to x \to y \to \ldots \to e,\ x \to y) \\
= w(p \to x \to y) - w(p \to x' \to e) \,,
\end{multline*}
where $x' \in \Cluster(x)$ is chosen to minimize $w(x' \to e)$.

In \LKs{}{}{}, we update both $x$ and $e$ such that the path $p \to x \to e \to r$ is minimized:
\begin{multline*}
\func{Gain}_\text{S}(b \to \ldots \to p \to x \to y \to \ldots \to r \to e,\ x \to y) = \\
w(p \to x \to y) + w(r \to e) - w(p \to x' \to e' \to r) \,,
\end{multline*}
where $x' \in \Cluster(x)$ and $e' \in \Cluster(e)$ are chosen to minimize $w(p \to x' \to e' \to r)$.

Observe that the most time-consuming part of \LK{} is the path optimization.  In case of the \LKs{}{}{} variation, the bottleneck is the gain evaluation function which takes $O(s^2)$ operations.  In order to reduce the number of gain evaluations in \LKs{}{}{}, we do not consider some edges $x \to y$.  In particular, we assume that the improvement is usually not larger than $w_{\min}(X, Y) - w_{\min}(X, E)$, where $X = \Cluster(x)$, $Y = \Cluster(y)$, $E = \Cluster(e)$ and $w_{\min}(A, B)$ is the weight of the shortest edge between some clusters $A$ and $B$:
$\displaystyle
w_{\min}(A, B) = \min_{a \in A, b \in B} w(a \to b) \,.
$
Obviously, all the values $w_{\min}(A, B)$ are precalculated.  Note that this speed-up heuristic is used only when $\var{depth} \ge \alpha$, see Algorithm~\ref{alg:lk_main}.

One can hardly speed up the $\func{Gain}$ function in \LKb{}{}{} or \LKc{}{}{}.

The \func{RearrangePath} function does some further cluster optimization in the \LKc{}{}{} variation:
\begin{multline*}
\func{RearrangePath}_\text{C}(b \to \ldots \to p \to x \to y \to \ldots \to e,\ x \to y) \\
= b \to \ldots \to p \to x' \to e \to \ldots \to y \,,
\end{multline*}
where $x' \in \Cluster(x)$ is chosen to minimize the weight $w(p \to x' \to e)$.  In \LKs{}{}{} it just repeats the optimization performed for the \func{Gain} evaluation:
\begin{multline*}
\func{RearrangePath}_\text{S}(b \to \ldots \to p \to x \to y \to \ldots \to r \to e,\ x \to y) \\
= b \to \ldots \to p \to x' \to e' \to r \to \ldots \to y \,,
\end{multline*}
where $x' \in \Cluster(x)$ and $e' \in \Cluster(e)$ are chosen to minimize $w(p \to x' \to e' \to r)$.

Every time we want to close up the path, both \LKc{}{}{} and \LKs{}{}{} try all the combinations of the end vertices to minimize the weight of the loop: 
\begin{multline*}
\func{CloseUp}_\text{C, S}(b \to p \to \ldots \to q \to e) = 
	b' \to p \to \ldots \to q \to e' \to b' :\ \\
	b' \in \Cluster(b), e' \in \Cluster(e) \text{ and } w(q \to e' \to b' \to p) \text{ is minimized} \,.
\end{multline*}

We also implemented the \LKc{}{}{co} and \LKs{}{}{co} variations such that \CO{} is applied every time a tour improvement is found (see Option~\ref{item:varyco} above): $\func{ImproveTour}(T) = \func{CO}(T)$.

\subsection{The Exact Variation}
\label{sec:exact}

Finally we propose the \heuristic{Exact} (\LKe{}{}) variation.  For every cluster ordering under consideration it finds the shortest path from the first to the last cluster (via all clusters in that order).  After closing up the path it always applies \CO{} (see Option~\ref{item:exact} above).  However, it explores the neighborhood much faster than a naive implementation would do.

The $\func{Gain}$ function for \LKe{}{} is defined as follows:
\begin{multline*}
\func{Gain}_\text{E}(b \to \ldots \to x \to y \to \ldots \to e,\ x \to y) = \\
w_\text{co}(b \to \ldots \to x \to e \to \ldots \to y) - w_\text{co}(b \to \ldots \to x \to y \to \ldots \to e) \,,
\end{multline*}
where $w_\text{co}(P)$ is the weight of the shortest path through the corresponding clusters: 
$$
w_\text{co}(x_1 \to x_2 \to \ldots \to x_m) = \min_{x'_i \in \Cluster(x_i), i = 1, \ldots, m} w(x'_1 \to x'_2 \to \ldots \to x'_m) \,.
$$
Note that $\func{ImprovePath}$ runs this function sequentially for every $x \to y \in P$\@.  In case of a naive implementation, it would take $O(m^2 s^2)$ operations.  Our implementation requires only $O(m s^3)$ operations but in practice it is much faster (almost $O(m s^2)$).  Also note that typically $m \gg s$.

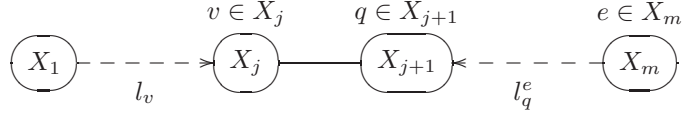
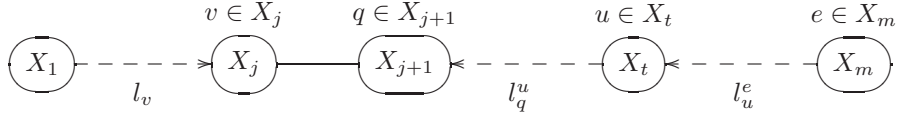
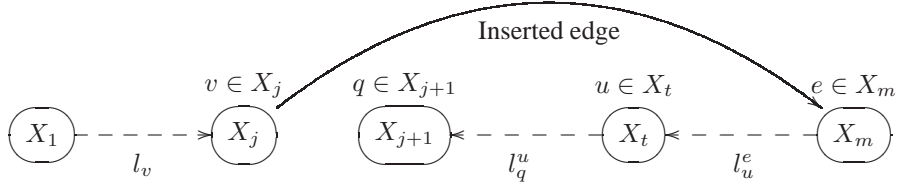
\begin{figure}[ht]
\centering  
\captionsetup{width=0.9\textwidth}
\subfloat[The original sequence of clusters $X_1$, $X_2$, \ldots, $X_m$.  The value $l_v$ denotes the shortest path from the cluster $X_1$ through $X_2$, $X_3$, \ldots, $X_{j-1}$ to the vertex $v \in X_j$.  It takes $O(|X_{j-1}| |X_j|)$ operations to calculate all $l_v$ for some $j$.  Value $l^e_q$ denotes the shortest path from the vertex $e \in X_m$ through $X_{m-1}$, $X_{m-2}$, \ldots, $X_{j+2}$ to the vertex $q \in X_{j+1}$.  It takes $O(|X_m| |X_{j+2}| |X_{j+1}|)$ operations to calculate all $l_q^e$ for some $j$.]{
\label{fig:original_exact_algorithm}
\xymatrix@R=0em@C=2em@L=0.1em{
	&
	&	v \in X_j
	&	q \in X_{j+1}
	&&e \in X_m
	\\	
		*++[F-:<1em>]{X_1} \ar@{-->}[rr]_*++{l_v}
	&&*++[F-:<1em>]{X_j}	\ar@{-}[r]
	&	*++[F-:<1em>]{X_{j+1}} \ar@{<--}[rr]_*++{l^e_q}
	&&*++[F-:<1em>]{X_m}
}}
\\[1em]
\subfloat[An improved algorithm.  Let cluster $X_t$ be the smallest cluster among $X_{j+2}$, $X_{j+3}$, \ldots, $X_m$.  To calculate all the shortest paths $l_q^u$ from $u \in X_t$ to $q \in X_{j+1}$ via $X_{t-1}$, $X_{t-2}$, \ldots, $X_{j+2}$, one needs $O(|X_t| |X_{j+2}| |X_{j+1}|)$ operations for some $j$, i.e., it is $|X_m| / |X_t|$ times faster than the straightforward calculations.  The values $l_u^e$ are calculated as previously, see Figure~\subref{fig:original_exact_algorithm}.]{
\label{fig:improved_exact_algorithm}
\xymatrix@R=0em@C=2em@L=0.1em{
	&
	&	v \in X_j	
	&	q \in X_{j+1}
	&&u \in X_t	
	&&e \in X_m	
	\\	
		*++[F-:<1em>]{X_1} \ar@{-->}[rr]_*++{l_v}
	&&*++[F-:<1em>]{X_j}	\ar@{-}[r]
	&	*++[F-:<1em>]{X_{j+1}}
	&&*++[F-:<1em>]{X_t} \ar@{-->}[ll]^*++{l^u_q}
	&&*++[F-:<1em>]{X_{m}} \ar@{-->}[ll]^*++{l^e_u}
}}
\\[1em]
\subfloat[The sequence of clusters after the local search move.  To find the shortest path from $X_1$ to $X_{j+1}$ via $X_2$, $X_3$, \ldots, $X_j$, $X_m$, $X_{m-1}$, \ldots, $X_{j+2}$, we need to find all the shortest paths $l'_e$ from $X_1$ to every $e \in X_m$ as $l'_e = \min_v \{ l_v + w(v \to e) \}$ in $O(s^2)$ operations, then find all the shortest paths $l'_u$ from $X_1$ to every $u \in X_t$ as $l'_u = \min_e \{ l'_e + l_u^e \}$ in $O(s^2)$ operations and, finally, find the whole shortest path $l'$ from $X_1$ to $X_{j+1}$ as $l' = \min_{u,q} \{ l'_u + l_q^u \}$ in $O(s^2)$ operations.]{
\label{fig:exact_path_calculation}
\xymatrix@R=0em@C=2em@L=0.1em{
	&
	&	v \in X_j	
	&	q \in X_{j+1}
	&&u \in X_t	
	&&e \in X_m	
	\\	
		*++[F-:<1em>]{X_1} \ar@{-->}[rr]_*++{l_v}
	&&*++[F-:<1em>]{X_j} \ar@/^5em/@{->}[rrrrr]_*++{\text{Inserted edge}}
	&	*++[F-:<1em>]{X_{j+1}}
	&&*++[F-:<1em>]{X_t} \ar@{-->}[ll]^*++{l^u_q}
	&&*++[F-:<1em>]{X_{m}} \ar@{-->}[ll]^*++{l^e_u}
}}

\captionsetup{width=\textwidth}
\caption{A straightforward and an enhanced implementations of the \LKe{}{} variation.}

\label{fig:exact_improve_chain}
\end{figure}

Our implementation proceeds as follows.  Let $X_1$, $X_2$, \ldots, $X_m$ be the sequence of clusters in the given path (see Figure~\ref{fig:original_exact_algorithm}).
Let $l_v$ be the length of the shortest path from $X_1$ to $v \in X_j$ through the cluster sequence $X_2$, $X_3$, \ldots, $X_{j-1}$:
$$
l_v = \min_{x_i \in X_i, i = 1, \ldots, j - 1} w(x_1 \to x_2 \to \ldots \to x_{j-1} \to v) \,.
$$
It takes $O(s^2 m)$ operations to calculate all $l_v$ using the algorithm for the shortest path in layered networks.

Let $l_q^e$ be the length of the shortest path from $e \in X_m$ to $q \in X_{j+1}$ through the cluster sequence $X_{m-1}$, $X_{m-2}$, \ldots, $X_{j+2}$:
$$
l_q^e = \min_{x_i \in X_i, i = j + 2, \ldots, m - 1} w(e \to x_{m-1} \to x_{m-2} \to \ldots \to x_{j+2} \to q) \,.
$$
It takes $O(s^3 m)$ operations to calculate all $l_q^e$ using the algorithm for the shortest path in layered networks.  

As a further improvement, we propose an algorithm to calculate $l_q^e$ which also takes $O(s^3 m)$ operations in the worst case but in practice it proceeds significantly faster.

Note that a disadvantage of a straightforward use of the shortest path algorithm to find $l_q^e$ is that its performance strongly depends on the size of $X_m$; indeed, the straightforward approach requires $|X_m| |X_{j+2}| |X_{j+1}|$ operations for every $j$.  Assume $|X_t| < |X_m|$ for some $t$, $j + 1 < t < m$, and we know the values $l_u^e$ for every $u \in X_t$ (see Figure~\ref{fig:improved_exact_algorithm}).  Now for every $j < t - 1$ we only need to calculate $l_q^u$, where $u \in X_t$ and $q \in X_{j+1}$.  This will take $|X_t| |X_{j+1}| |X_j|$ operations for every $j$, i.e, it is $|X_m| / |X_t|$ times faster than the straightforward approach.
A formal procedure is shown in Algorithm~\ref{alg:shortestpaths}.

\begin{algorithm}[ht]
\caption{Calculation of the shortest paths $l_u^e$ and $l_q^u$ for \LKe{}{}.}
\label{alg:shortestpaths}

\begin{algorithmic}
\REQUIRE The sequence of clusters $X_1$, $X_2$, \ldots, $X_m$.

\FOR {every $e \in X_m$ and every $q \in X_{m-1}$}
	\STATE $l_q^e \gets w(e \to q)$.
\ENDFOR

\STATE $Y \gets X_m$.

\FOR {$j \gets m - 3, m - 4, \ldots, 1$}
	\IF {$|X_{j+2}| < |Y|$}
		\IF {$Y \neq X_m$}
			\FOR {every $e \in X_m$ and every $u \in X_{j+2}$}
				\STATE $l_u^e \gets \min_{y \in Y} \{ l_y^e + l_u^y \}$.
			\ENDFOR
		\ENDIF
		\STATE $Y \gets X_{j+2}$.
	\ENDIF

	\FOR {every $y \in Y$ and every $q \in X_{j+1}$}
		\STATE $l_q^y \gets \min_{u \in X_j} \{ l_u^y + w(u \to q) \}$.
	\ENDFOR
\ENDFOR
\end{algorithmic}
\end{algorithm}

Having all $l_v$, $l_u^e$ and $l_q^u$, where $v \in X_j$, $q \in X_{j+1}$, $e \in X_m$ and $u \in X_t$, $j + 1 < t < m$, one can find the shortest path through all the clusters $X_1$, $X_2$, \ldots, $X_j$, $X_m$, $X_{m-1}$, \ldots, $X_{j+1}$ in $O(s^2)$ time, see Algorithm~\ref{alg:shortestpath} and Figure~\ref{fig:exact_path_calculation}.

\begin{algorithm}[ht]
\caption{Calculation of the whole shortest path for \LKe{}{}.}
\label{alg:shortestpath}

\begin{algorithmic}
\REQUIRE The index $j$.
\REQUIRE The values $l_v$, $l_u^e$ and $l_q^u$, where $v \in X_j$, $q \in X_{j+1}$, $e \in X_m$ and $u \in X_t$, $j + 1 < t \le m$.

\STATE Calculate $l_e \gets \min_{v \in X_j} l_v + w(v \to e)$ for every $e \in X_m$.
\IF {$t < m$}
	\STATE Calculate $l_u \gets \min_{e \in X_m} l_e + l_u^e$ for every $u \in X_t$.
	\STATE Calculate $l_q \gets \min_{u \in X_t} l_u + l_q^u$ for every $q \in X_{j+1}$.
\ELSE
	\STATE Calculate $l_q \gets \min_{e \in X_m} l_e + l_q^e$ for every $q \in X_{j+1}$.
\ENDIF
	\RETURN $\min_{q \in X_{j+1}} l_q$.
\end{algorithmic}
\end{algorithm}

In our experiments this speed-up heuristic decreased the running time of the \LKe{}{} algorithm by 30\% to 50\%.


The \func{RearrangePath} function for \LKe{}{} replaces the edge $x \to y$ with $x \to e$ and optimizes the vertices in the path:
\begin{multline*}
\func{RearrangePath}_\text{E}(b \to \ldots \to x \to y \to \ldots \to e) = b' \to \ldots \to x' \to y' \to \ldots \to e' \,,
\end{multline*}
where all the vertices are selected to minimize the weight of the resulting path.  The \func{CloseUp} function for \LKe{}{} simply applies \CO{} to the tour:
$$
\func{CloseUp}_\text{E}(b \to \ldots \to e) = \func{CO}(b \to \ldots \to e \to b) \,.
$$

Observe that, unlike other adaptations of the original \LKtsp{} heuristic, \heuristic{Exact} is naturally suitable for asymmetric instances.

\bigskip

Note that another approach to implement the \CO{} algorithm is proposed by \citet{Pop2007}.  It is based on an integer formulation of the GTSP; a more general case is studied in~\citep{Pop2006}.  However, we believe that the dynamic programming approach enhanced by the improvements discussed above is more efficient in our case.

\subsection{The Gain Function}
\label{sec:gain}

The gain is a measure of a path improvement.  It is used to find the best path improvement and to decide whether this improvement should be accepted.  To decide this, we use a boolean function $\func{GainIsAcceptable}(P,\ x \to y)$.  This function greatly influences the performance of the whole algorithm.  We propose four different implementations of $\func{GainIsAcceptable}(P,\ x \to y)$ in order to find the most efficient ones.  For the notation, see Algorithm~\ref{alg:lk_main}.
\begin{enumerate}
	\item \label{item:gain1} $\func{GainIsAcceptable}(P,\ x \to y) = w(P) < w(P_o)$, i.e., the function accepts any changes while the path is shorter than the original one.
	
	\item \label{item:gain2} $\func{GainIsAcceptable}(P,\ x \to y) = w(P) + \frac{w(T)}{m} < w(T)$, i.e., it is assumed that an edge of an average weight $\frac{w(T)}{m}$ will close up the path.
	
	\item \label{item:gain3} $\func{GainIsAcceptable}(P,\ x \to y) = w(P) + w(x \to y) < w(T)$, i.e., the last removed edge is `restored' for the gain evaluation.  Note that the weight of the edge $x \to y$ cannot be obtained correctly in \LKe{}{}.  Instead of $w(x \to y)$ we use the weight $w_{\min}(X, Y)$ of the shortest edge between $X = \Cluster(x)$ and $Y = \Cluster(y)$.

	\item \label{item:gain4} $\func{GainIsAcceptable}(P,\ x \to y) = w(P) < w(T)$, i.e., the obtained path has to be shorter than the original tour.  In other words, the weight of the `close up edge' is assumed to be 0.  Unlike the first three implementations, this one is optimistic and, hence, yields deeper search trees.  This takes more time but also improves the solution quality.

	\item \label{item:gain5} $\func{GainIsAcceptable}(P,\ x \to y) = w(P) + \frac{w(T)}{2m} < w(T)$, i.e., it is assumed that an edge of a half of an average weight will close up the path.  It is a mixture of Options~\ref{item:gain2} and~\ref{item:gain4}.
\end{enumerate}

\section{Experiments}
\label{sec:experiments}

In order to select the most successful variations of the proposed heuristic and to prove its efficiency, we conducted a set of computational experiments.  

Our test bed includes several TSP instances taken from TSPLIB~\citep{TSPLIB} converted into the GTSP by the standard clustering procedure of \citet{Fischetti1997} (the same approach is widely used in the literature, see, e.g., \citep{GK_GTSP_GA_2008,Silberholz2007,Snyder2000,Tasgetiren2007}).  Like \citet{Bontoux2009}, \citet{GK_GTSP_GA_2008}, and \citet{Silberholz2007}, we do not consider any instances with less than 10 or more than 217 clusters (in other papers the bounds are stricter).

Every instance name consists of three parts: `$m$ $t$ $n$', where $m$ is the number of clusters, $t$ is the type of the original TSP instance (see~\citep{TSPLIB} for details) and $n$ is the number of vertices.

Observe that the optimal solutions are known only for some instances with up to 89 clusters~\citep{Fischetti1997}.  For the rest of the instances we use the best known solutions, see~\citep{Bontoux2009,GK_GTSP_GA_2008,Silberholz2007}.

The following heuristics were included in the experiments:
\begin{enumerate}
	\item The \heuristic{Basic} variations, i.e., \LKb{$x$}{$\alpha$}{} and \LKb{$x$}{$\alpha$}{\,co}, where $\alpha \in \{ 2, 3, 4 \}$ and $x \in \{ 1, 2, 3, 4, 5 \}$ define the backtracking depth and the gain acceptance strategy, respectively.  The letters `co' in the superscript mean that the \CO{} algorithm is applied every time a tour improvement is found (for details see Section~\ref{sec:ls_adaptation}).
	
	\item The \heuristic{Closest} variations, i.e., \LKc{$x$}{$\alpha$}{} and \LKc{$x$}{$\alpha$}{\,co}, where $\alpha \in \{ 2, 3, 4 \}$ and $x \in \{ 1, 2, 3, 4, 5 \}$.

	\item The \heuristic{Shortest} variations, i.e., \LKs{$x$}{$\alpha$}{} and \LKs{$x$}{$\alpha$}{\,co}, where $\alpha \in \{ 2, 3, 4 \}$ and $x \in \{ 1, 2, 3, 4, 5 \}$.

	\item The \heuristic{Exact} variations, i.e., \LKe{$x$}{$\alpha$}, where $\alpha \in \{ 1, 2, 3 \}$ and $x \in \{ 1, 2, 3, 4, 5 \}$.

	\item Adaptations of the 2-opt (\optshort{2}{}{}) and 3-opt (\optshort{3}{}{}) local searches according to Section~\ref{sec:ls_adaptation}.
	
	\item A state-of-the-art memetic algorithm \MA{} by~\citet{GK_GTSP_GA_2008}.
\end{enumerate}

Observe that \MA{} dominates all other GTSP metaheuristics known from the literature.  In particular, \citet{GK_GTSP_GA_2008} compare it to the heuristics proposed by \citet{Silberholz2007}, \citet{Snyder2000} and \citet{Tasgetiren2007}, and it appears that \MA{} dominates all these algorithms in every experiment with respect to both solution quality and running time.  Similarly, one can see that it dominates two more recent algorithms by \citet{Bontoux2009} and \citet{Tasgetiren2010} in every experiment.  Note that the running times of all these algorithms were normalized according to the computational platforms used to evaluate the algorithms.  Hence, we do not include the results of the other metaheuristics in our comparison.

In order to generate the starting tour for the local search procedures, we use a simplified Nearest Neighbour construction heuristic (\NN)\@.  Unlike proposed by \citet{Noon1988}, our algorithm tries only one starting vertex.  Trying every vertex as a starting one significantly slows down the heuristic and usually does not improve the solutions of the local searches.  Note that in what follows the running time of a local search includes the running time of the construction heuristic.

All the heuristics are implemented in Visual C++\@.  The evaluation platform is based on an Intel Core i7 2.67~GHz processor.

The experimental results are presented in two forms.  The first form is a fair competition of all the heuristics joined in one table.  The second form is a set of standard tables reporting solution quality and running time of the most successful heuristics.

\subsection{Heuristics Competition}
\label{sec:fair_competition}

Many researchers face the problem of a fair comparison of several heuristics.  Indeed, every experiment result consist of at least two parameters: solution error and running time.  It is a trade-off between the speed and the quality, and both quick (and low-quality) and slow (and high-quality) heuristics are of interest.  A heuristic should only be considered as useless if it is \emph{dominated} by another heuristic, i.e., it is both slower and yields solutions of a lower quality.

Hence, one can clearly separate a set of successful from a set of dominated heuristics.  However, this only works for a single experiment.  If the experiment is conducted for several test instances, the comparison becomes not obvious.  Indeed, a heuristic may be successful in one experiment and unsuccessful in another one.  A natural solution of this problem is to use averages but if the results vary a lot for different instances this approach may be incorrect.

In a fair competition, one should compare heuristics which have similar running times.  For every time $\tau_i \in $ \{0.02 s, 0.05 s, 0.1 s, 0.2 s, \ldots, 50 s\} we compare solution quality of all the heuristics which were able to solve an instance in less than $\tau_i$.  In order to further reduce the size of the table and to smooth out the experimental results, we additionally group similar instances together and report only the average values for each group.

Moreover, we repeat every experiment 10 times.  It requires some extra effort to ensure that an algorithm $H$ proceeds differently in different runs, i.e., $H^i(I) \neq H^j(I)$ in general case, where $i$ and $j$ are the run numbers.  For \MA$^r$ the run number $r$ is the random generator seed value.  In \NN$^{r}$, we start the tour construction from the vertex $C_{r,1}$, i.e., from the first vertex of the $r$th cluster of the instance.  This also affects all the local searches since they start from the \NN$^r$ solutions.

Finally we get Table~\ref{tab:competition}.  Roughly speaking, every cell of this table reports the most successful heuristics for a given range of instances and being given some limited time.  More formally, let $\tau = \{ \tau_1, \tau_2, \ldots \}$ be a set of predefined time limits.  Let $\mathcal{I} = \{ \mathcal{I}_1, \mathcal{I}_2, \ldots \}$ be a set of predefined instance groups such that all instances in every $\mathcal{I}_j$ have similar difficulty.  Let $\mathcal{H}$ be a set of all heuristics included in the competition.  $H(I)_\text{time}$ and $H(I)_\text{error}$ are the running time and the relative solution error, respectively, of the heuristic $H \in \mathcal{H}$ for the instance $I \in \mathcal{I}$:
$$
H(I)_\text{error} = \frac{w(H(I)) - w(I_\text{best})}{w(I_\text{best})} \,,
$$
where $I_\text{best}$ is the optimal or the best known solution for the instance $I$.  $H(\mathcal{I}_j)_\text{time}$ and $H(\mathcal{I}_j)_\text{error}$ denote the corresponding values averaged for all the instances $I \in \mathcal{I}_j$ and all $r \in \{ 1, 2, \ldots, 10 \}$.

For every cell $i,j$ we define a winner heuristic $\var{Winner}_{i,j} \in \mathcal{H}$ as follows:
\begin{enumerate}
	\item $\var{Winner}_{i,j}^r(I)_\text{time} \le \tau_i$ for every instance $I \in \mathcal{I}_j$ and every $r \in \{ 1, 2, \ldots, 10 \}$.
	\item $\var{Winner}_{i,j}(\mathcal{I}_j)_\text{error} < \var{Winner}_{i-1,j}(\mathcal{I}_j)_\text{error}$ (it is only applicable if $i > 1$).
	\item If several heuristics meet the conditions above, we choose the one with the smallest $H_{i,j}(\mathcal{I}_j)_\text{error}$. 
	\item If several heuristics meet the conditions above and have the same solution quality, we choose the one with the smallest $H_{i,j}(\mathcal{I}_j)_\text{time}$.
\end{enumerate}

Apart from the winner, every cell contains all the heuristics $H \in \mathcal{H}$ meeting the following conditions:
\begin{enumerate}
	\item $H^r(I)_\text{time} \le \tau_i$ for every instance $I \in \mathcal{I}_j$ and every $r \in \{ 1, 2, \ldots, 10 \}$.
	\item $H(\mathcal{I}_j)_\text{error} < \var{Winner}_{i-1,j}(\mathcal{I}_j)_\text{error}$ (it is only applicable if $i > 1$).
	\item $H(\mathcal{I}_j)_\text{error} \le 1.1 \cdot Winner_{i,j}(\mathcal{I}_j)_\text{error}$.
	\item $H(\mathcal{I}_j)_\text{time} \le 1.2 \cdot Winner_{i,j}(\mathcal{I}_j)_\text{time}$.
\end{enumerate}

Since \LK{} is a powerful heuristic, we did not consider any instances with less than 30 clusters in this competition.  Note that all the smaller instances are relatively easy to solve, e.g., \MA{} was able to solve all of them to optimality in our experiments, and it took only about 30~ms on average, and for \LKs{5}{2}{co} it takes, on average, less than 0.5~ms to get 0.3\% error, see Table~\ref{tab:detailed_small}.  

We use the following groups $\mathcal{I}_j$ of instances:\\
Tiniest: 	
	\instance{30ch150},
	\instance{30kroA150},
	\instance{30kroB150},
	\instance{31pr152},
	\instance{32u159} and
	\instance{39rat195}.
\\
Tiny:
	\instance{40kroa200},
	\instance{40krob200},
	\instance{41gr202},
	\instance{45ts225},
	\instance{45tsp225} and
	\instance{46pr226}.
\\
Small:
	\instance{46gr229},
	\instance{53gil262},
	\instance{56a280},
	\instance{60pr299} and
	\instance{64lin318}.
\\	
Moderate:
	\instance{80rd400},
	\instance{84fl417},
	\instance{87gr431},
	\instance{88pr439} and
	\instance{89pcb442}.
\\
Large:
	\instance{99d493},
	\instance{107att532},
	\instance{107ali535},
	\instance{113pa561},
	\instance{115u574} and
	\instance{115rat575}.
\\
Huge:
	\instance{132d657},
	\instance{134gr666},
	\instance{145u724} and
	\instance{157rat783}.
\\
Giant:
	\instance{200dsj1000},
	\instance{201pr1002},
	\instance{212u1060} and
	\instance{217vm1084}.

Note that the instances \instance{35si175}, \instance{36brg180}, \instance{40d198}, \instance{53pr264}, \instance{107si535}, \instance{131p654} and \instance{207si1032} are excluded from this competition since they are significantly harder to solve than the other instances of the corresponding groups.  This is discussed in Section~\ref{sec:detailed_results} and the results for these instances are included in Tables~\ref{tab:quality} and~\ref{tab:time}.

\begin{table}[!ht] \centering
\footnotesize
\caption{The fair competition.  Every cell reports the most successful heuristics being given some limited time (see the first column) for a given range of instances (see the header).  Every heuristic is provided with the average relative solution error in percent.  To make the table easier to read, all the \LKb{}{}{} and \LKe{}{} adaptations of \LK{} are selected with bold font.  All the cells where the dominating heuristic is \LKc{}{}{} or \LKs{}{}{} are highlighted with grey background.}
\label{tab:competition}
\begin{tabular}{r @{\hspace{0.5em}} *{7}{@{\hspace{0.5em}} c}}
\toprule

&Tiniest&Tiny&Small&Moderate&Large&Huge&Giant\\
\cmidrule(){1-8}

$\le 2$ ms&\cellcolor{grey}\begin{tabular}{p{1.7em}@{}l}{\LKs{4}{2}{co}} & \phantom{1}1.2\\ {\LKs{5}{2}{co}} & \phantom{1}1.2\\ {\LKc{5}{2}{co}} & \phantom{1}1.3\\ \end{tabular}&\cellcolor{grey}\begin{tabular}{p{1.7em}@{}l}{\LKc{5}{2}{co}} & \phantom{1}1.0\\ \end{tabular}&\cellcolor{grey}\begin{tabular}{p{1.7em}@{}l}{\LKc{1}{2}{co}} & \phantom{1}3.5\\ \end{tabular}&\begin{tabular}{p{1.7em}@{}l}\textbf{\LKb{5}{2}{co}} & \phantom{1}6.1\\ \textbf{\LKb{2}{2}{co}} & \phantom{1}6.1\\ \textbf{\LKb{1}{4}{co}} & \phantom{1}6.3\\ \textbf{\LKb{3}{2}{co}} & \phantom{1}6.5\\ \end{tabular}&\begin{tabular}{p{1.7em}@{}l}\textbf{\LKb{1}{2}{co}} & \phantom{1}7.8\\ \end{tabular}&\begin{tabular}{p{1.7em}@{}l}{\optshort{2}{B}{co}} & 13.4\\ \end{tabular}&\begin{tabular}{p{1.7em}@{}l}{\optshort{2}{B}{}} & 22.7\\ \end{tabular}\\
\cmidrule(){1-8}

$\le 5$ ms&\cellcolor{grey}\begin{tabular}{p{1.7em}@{}l}{\LKs{5}{3}{co}} & \phantom{1}0.0\\ \end{tabular}&\cellcolor{grey}\begin{tabular}{p{1.7em}@{}l}{\LKc{5}{3}{co}} & \phantom{1}0.5\\ \end{tabular}&\cellcolor{grey}\begin{tabular}{p{1.7em}@{}l}{\LKc{5}{3}{co}} & \phantom{1}1.2\\ {\LKs{5}{2}{co}} & \phantom{1}1.2\\ \end{tabular}&\cellcolor{grey}\begin{tabular}{p{1.7em}@{}l}{\LKc{5}{2}{co}} & \phantom{1}2.4\\ \end{tabular}&\begin{tabular}{p{1.7em}@{}l}\textbf{\LKb{1}{4}{co}} & \phantom{1}7.2\\ \textbf{\LKb{2}{2}{co}} & \phantom{1}7.3\\ \end{tabular}&\begin{tabular}{p{1.7em}@{}l}\textbf{\LKb{5}{2}{co}} & \phantom{1}9.5\\ \textbf{\LKb{1}{3}{co}} & \phantom{1}9.6\\ \textbf{\LKb{3}{2}{co}} & 10.1\\ \textbf{\LKb{2}{2}{co}} & 10.3\\ \end{tabular}&\begin{tabular}{p{1.7em}@{}l}{\optshort{2}{B}{co}} & 14.3\\ \end{tabular}\\
\cmidrule(){1-8}

$\le 10$ ms&\cellcolor{grey}---&\cellcolor{grey}---&\cellcolor{grey}\begin{tabular}{p{1.7em}@{}l}{\LKc{5}{4}{co}} & \phantom{1}0.8\\ \end{tabular}&\cellcolor{grey}\begin{tabular}{p{1.7em}@{}l}{\LKc{4}{2}{co}} & \phantom{1}1.3\\ \end{tabular}&\cellcolor{grey}\begin{tabular}{p{1.7em}@{}l}{\LKc{5}{2}{co}} & \phantom{1}2.9\\ \end{tabular}&\cellcolor{grey}\begin{tabular}{p{1.7em}@{}l}{\LKc{1}{2}{co}} & \phantom{1}6.1\\ {\LKc{2}{2}{co}} & \phantom{1}6.3\\ \end{tabular}&\begin{tabular}{p{1.7em}@{}l}\textbf{\LKb{3}{2}{co}} & \phantom{1}7.9\\ \end{tabular}\\
\cmidrule(){1-8}

$\le 20$ ms&\cellcolor{grey}---&\cellcolor{grey}\begin{tabular}{p{1.7em}@{}l}{\LKs{4}{3}{co}} & \phantom{1}0.5\\ \end{tabular}&\cellcolor{grey}\begin{tabular}{p{1.7em}@{}l}{\LKs{5}{3}{co}} & \phantom{1}0.4\\ {\LKs{2}{4}{co}} & \phantom{1}0.5\\ \end{tabular}&\cellcolor{grey}\begin{tabular}{p{1.7em}@{}l}{\LKc{5}{3}{co}} & \phantom{1}1.3\\ \end{tabular}&\cellcolor{grey}\begin{tabular}{p{1.7em}@{}l}{\LKc{4}{2}{co}} & \phantom{1}2.4\\ \end{tabular}&\cellcolor{grey}\begin{tabular}{p{1.7em}@{}l}{\LKc{5}{2}{co}} & \phantom{1}4.0\\ \end{tabular}&---\\
\cmidrule(){1-8}

$\le 50$ ms&\cellcolor{grey}---&\cellcolor{grey}\begin{tabular}{p{1.7em}@{}l}{\LKs{4}{4}{}} & \phantom{1}0.2\\ \end{tabular}&\cellcolor{grey}\begin{tabular}{p{1.7em}@{}l}{\LKs{5}{4}{}} & \phantom{1}0.2\\ \end{tabular}&\cellcolor{grey}\begin{tabular}{p{1.7em}@{}l}{\LKs{2}{4}{co}} & \phantom{1}1.1\\ \end{tabular}&\cellcolor{grey}\begin{tabular}{p{1.7em}@{}l}{\LKs{1}{3}{}} & \phantom{1}2.2\\ {\LKs{4}{2}{co}} & \phantom{1}2.2\\ \end{tabular}&\cellcolor{grey}\begin{tabular}{p{1.7em}@{}l}{\LKs{5}{2}{co}} & \phantom{1}2.9\\ {\LKc{5}{3}{co}} & \phantom{1}3.0\\ \end{tabular}&\cellcolor{grey}\begin{tabular}{p{1.7em}@{}l}{\LKc{2}{2}{co}} & \phantom{1}4.0\\ \end{tabular}\\
\cmidrule(){1-8}

$\le 0.1$ s&\cellcolor{grey}---&\cellcolor{grey}\begin{tabular}{p{1.7em}@{}l}{\LKs{4}{4}{co}} & \phantom{1}0.2\\ \end{tabular}&\cellcolor{grey}\begin{tabular}{p{1.7em}@{}l}{\LKs{4}{4}{co}} & \phantom{1}0.0\\ \end{tabular}&\cellcolor{grey}---&\cellcolor{grey}\begin{tabular}{p{1.7em}@{}l}{\LKc{4}{3}{co}} & \phantom{1}1.0\\ \end{tabular}&\cellcolor{grey}\begin{tabular}{p{1.7em}@{}l}{\LKc{4}{3}{co}} & \phantom{1}1.7\\ \end{tabular}&\cellcolor{grey}\begin{tabular}{p{1.7em}@{}l}{\LKs{2}{2}{co}} & \phantom{1}3.0\\ \end{tabular}\\
\cmidrule(){1-8}

$\le 0.2$ s&\cellcolor{grey}---&\cellcolor{grey}---&\cellcolor{grey}---&\begin{tabular}{p{1.7em}@{}l}\textbf{\LKe{3}{2}{}} & \phantom{1}0.6\\ \end{tabular}&\cellcolor{grey}---&\cellcolor{grey}---&\cellcolor{grey}\begin{tabular}{p{1.7em}@{}l}{\LKs{4}{2}{co}} & \phantom{1}1.9\\ \end{tabular}\\
\cmidrule(){1-8}

$\le 0.5$ s&\cellcolor{grey}---&\begin{tabular}{p{1.7em}@{}l}{\MA} & \phantom{1}0.0\\ \end{tabular}&\cellcolor{grey}---&---&\cellcolor{grey}---&\cellcolor{grey}\begin{tabular}{p{1.7em}@{}l}{\LKs{4}{3}{co}} & \phantom{1}1.2\\ \end{tabular}&\cellcolor{grey}---\\
\cmidrule(){1-8}

$\le 1$ s&\cellcolor{grey}---&---&\cellcolor{grey}---&\begin{tabular}{p{1.7em}@{}l}\textbf{\LKe{5}{3}{}} & \phantom{1}0.4\\ \end{tabular}&\cellcolor{grey}---&\begin{tabular}{p{1.7em}@{}l}\textbf{\LKe{2}{3}{}} & \phantom{1}1.0\\ \end{tabular}&\cellcolor{grey}\begin{tabular}{p{1.7em}@{}l}{\LKs{5}{3}{co}} & \phantom{1}1.2\\ \end{tabular}\\
\cmidrule(){1-8}

$\le 2$ s&\cellcolor{grey}---&---&\cellcolor{grey}---&---&\cellcolor{grey}---&\cellcolor{grey}\begin{tabular}{p{1.7em}@{}l}{\LKs{4}{4}{}} & \phantom{1}1.0\\ \end{tabular}&\cellcolor{grey}---\\
\cmidrule(){1-8}

$\le 5$ s&\cellcolor{grey}---&---&\cellcolor{grey}---&\begin{tabular}{p{1.7em}@{}l}{\MA} & \phantom{1}0.0\\ \end{tabular}&\begin{tabular}{p{1.7em}@{}l}\textbf{\LKe{5}{3}{}} & \phantom{1}0.8\\ \end{tabular}&\begin{tabular}{p{1.7em}@{}l}\textbf{\LKe{3}{3}{}} & \phantom{1}0.8\\ \end{tabular}&\cellcolor{grey}---\\
\cmidrule(){1-8}

$\le 10$ s&\cellcolor{grey}---&---&\cellcolor{grey}---&---&\begin{tabular}{p{1.7em}@{}l}{\MA} & \phantom{1}0.0\\ \end{tabular}&---&\cellcolor{grey}---\\
\cmidrule(){1-8}

$\le 20$ s&\cellcolor{grey}---&---&\cellcolor{grey}---&---&---&\begin{tabular}{p{1.7em}@{}l}{\MA} & \phantom{1}0.1\\ \end{tabular}&\cellcolor{grey}---\\
\cmidrule(){1-8}

$\le 50$ s&\cellcolor{grey}---&---&\cellcolor{grey}---&---&---&---&\begin{tabular}{p{1.7em}@{}l}{\MA} & \phantom{1}0.2\\ \end{tabular}\\

\bottomrule

\end{tabular}
\end{table}

One can see from Table~\ref{tab:competition} that there is a clear tendency: the proposed Lin-Kernighan adaptation outperforms all the other heuristics in a wide range of trade-offs between solution quality and running time.  Only the state-of-the-art memetic algorithm \MA{} is able to beat \LK{} being given large time.  There are several occurrences of \opt{2}{}{} in the upper right corner (i.e., for Huge and Giant instances and less than 5~ms time) but this is because this time is too small for even the most basic variations of \LK{}\@.  Note that \optshort{2}{B}{} and \optshort{2}{B}{co} denote the \opt{2}{}{} local search adapted for the GTSP according to Options~\ref{item:plain} and~\ref{item:applyco}, respectively, see Section~\ref{sec:ls_adaptation}.

Clearly, the most important parameter of \LK{} is its variation, and each of the four variations (\heuristic{Basic}, \heuristic{Closest}, \heuristic{Shortest} and \heuristic{Exact}) is successful in a certain running time range.  \LKb{}{}{} wins the competition for small running times.  For the middle range of running times one should choose \LKc{}{}{} or \LKs{}{}{}.  The \LKe{}{} variation wins only in a small range of times; having more time, one should choose the memetic algorithm \MA{}.

Here are some tendencies with regards to the rest of the \LK{} parameters:
\begin{itemize}
	\item It is usually beneficial to apply \CO{} every time a tour improvement is found.
	\item The most successful gain acceptance options are~\ref{item:gain4} and~\ref{item:gain5} (see Section~\ref{sec:gain}).
	\item The larger the backtracking depth $\alpha$, the better the solutions.  However, it is an expensive way to improve the solutions; one should normally keep $\alpha \in \{ 2, 3, 4 \}$.
\end{itemize}

Table~\ref{tab:competition}, however, does not make it clear what parameters one should use in practice.  In order to give some advice, we calculated the distances $d(H)$ between each heuristic $H \in \mathcal{H}$ and the winner algorithms.  For every column $j$ of Table~\ref{tab:competition} we calculated $d_j(H)$:
$$
d_j(H) = \frac{H(\mathcal{I}_j)_\text{error} - \var{Winner}_{i,j}(\mathcal{I}_j)_\text{error}}{\var{Winner}_{i,j}(\mathcal{I}_j)_\text{error}} \,,
$$
where $i$ is minimized such that $H^r(I)_\text{time} \le \tau_i$ for every $I \in \mathcal{I}_j$ and $r \in \{ 1, 2, \ldots, 10 \}$.  Then $d_j(H)$ were averaged for all $j$ to get the required distance: $d(H) = \overline{d_j(H)}$\@.  The list of the heuristics $H$ with the smallest distances $d(H)$ is presented in Table~\ref{tab:bestheuristics}.  In fact, we added \optshort{2}{B}{co}, \LKb{2}{2}{co} and \LKe{4}{2}{} to this list only to fill the gaps.  Every heuristic $H$ in Table~\ref{tab:bestheuristics} is also provided with the average running time $T(H)$, in \% of \MA{} running time:
\begin{multline*}
T(H) = \overline{T(H, I, r)} \text{ is averaged for all the instances $I \in \mathcal{I}$ and all $r \in \{1, 2, \ldots, 10 \}$} \,, \\
\text{where } T(H, I, r) = \frac{H^r(I)_\text{time}}{\func{MA}(I)_\text{time}} \\
\text{and } \func{MA}(I)_\text{time} = \overline{\func{MA}^r(I)_\text{time}} \text{ is averaged for all $r \in \{1, 2, \ldots, 10 \}$} \,.
\end{multline*}

\begin{table}[htb]
\caption{The list of the most successful heuristics.  The heuristics $H$ are ordered according to their running times, from the fastest to the slowest ones.  \optshort{2}{B}{co} denotes the \opt{2}{}{} local search adapted for the GTSP according to Option~\ref{item:applyco}, see Section~\ref{sec:ls_adaptation}.}
\label{tab:bestheuristics}
\begin{center}
\begin{tabular}{l @{\qquad} r @{\qquad} r}
\toprule
$H$ & $d(H)$, \% & Time, \% of \MA{} time \\
\cmidrule{1-3}

\optshort{2}{B}{co} & 44 & 0.04\\
\LKb{2}{2}{co} & 34 & 0.10\\
\LKc{5}{2}{co} & 12 & 0.40\\
\LKs{5}{2}{co} & 19 & 0.97\\
\LKs{5}{3}{co} & 19 & 2.53\\
\LKs{5}{4}{co} & 35 & 8.70\\
\LKs{4}{4}{co} & 32 & 15.34\\
\LKe{4}{2}{} & 56 & 43.62\\
\MA & 0 & 100.00\\

\bottomrule
\end{tabular}
\end{center}
\end{table}

\subsection{Detailed Data For Selected Heuristics}
\label{sec:detailed_results}

In this section we provide the detailed information on the experimental results for the most successful heuristics, see Section~\ref{sec:fair_competition}.  Tables~\ref{tab:detailed_small}, \ref{tab:quality} and~\ref{tab:time} include the following information:
\begin{itemize}
	\item The `Instance' column contains the instance name as described above.

	\item The `Best' column contains the best known or optimal \citep{Fischetti1997} objective values of the test instances.
	
	\item The rest of the columns correspond to different heuristics and report either relative solution error or running time in milliseconds.  Every value is averaged for ten runs, see Section~\ref{sec:fair_competition} for details.
	
	\item The `Average' row reports the averages for all the instances in the table.

	\item The `Light avg' row reports the averages for all the instances used in Section~\ref{sec:fair_competition}.

	\item Similarly, the `Heavy avg' row reports the averages for all the instances ($m \ge 30$) excluded from the competition in Section~\ref{sec:fair_competition}.
\end{itemize}

All the small instances ($m < 30$) are separated from the rest of the test bed to Table~\ref{tab:detailed_small}.  One can see that all these instances are relatively easy to solve; in fact several heuristics are able to solve all or almost all of them to optimality in every run and it takes only a small fraction of a second.  A useful observation is that \LKe{4}{2} solves all the instances with up to 20 clusters to optimality, and in this range \LKe{4}{2} is significantly faster than \MA{}.

As regards the larger instances ($m \ge 30$), it is worth noting that there exist several `heavy' instances among them: \instance{35si175}, \instance{36brg180}, \instance{40d198}, \instance{53pr264}, \instance{107si535}, \instance{131p654} and \instance{207si1032}\@.  Some heuristics perform extremely slowly for these instances: the running time of \LKs{5}{3}{co}, \LKs{5}{4}{co}, \LKs{4}{4}{co} and \LKe{4}{2} is 3 to 500 times larger for every `heavy` instance than it is for the other instances of a similar size.  Other \LK{} variations are also affected, though, this mostly relates to the ones which use the `optimistic' gain acceptance functions (Options~\ref{item:gain4} and~\ref{item:gain5}), see Section~\ref{sec:gain}.

Our analysis has shown that all of these instances have an unusual weight distribution.  In particular, all these instances have enormous number of `heavy' edges, i.e., the the weights which are close to the maximum weight in the instance, prevail over the smaller weights.  Recall that \LK{} bases on the assumption that a randomly selected edge will probably have a `good' weight.  Then we can optimize a path in the hope to find a good option to close it up later.  However, the probability to find a `good' edge is low in a `heavy' instance.  Hence, the termination condition \func{GainIsAcceptable} does not usually stop the search though a few tour improvements can be found.  This, obviously, slows down the algorithm.

Note that a similar result was obtained by \citet{GK_MAP_LS_2010} for the adaptation of the Lin-Kernighan heuristic for the Multidimensional Assignment Problem.

Observe that such `unfortunate' instances can be easily detected before the algorithm's run.  Observe also that even the fast heuristics yield relatively good solutions for these instances (see Tables~\ref{tab:quality} and~\ref{tab:time}).  Hence, one can use a lighter heuristic to get a reasonable solution quality in a reasonable time in this case.

\section{Conclusion}
\label{sec:conclusion}

The Lin-Kernighan heuristic is known to be a very successful TSP heuristic.  In this paper we present a number of adaptations of Lin-Kernighan for the GTSP\@.  Several approaches to adaptation of a TSP local search for the GTSP are discussed and the best ones are selected and applied to the Lin-Kernighan heuristic.  The experimental evaluation confirms the success of these approaches and proves that the proposed adaptations reproduce the efficiency of the original TSP heuristic.

Based on the experimental results, we selected the most successful Lin-Kernighan adaptations for different solution quality/running time requirements.  Only for the very small running times (5 ms or less) and huge instances (132 clusters and more) our heuristic is outperformed by some very basic local searches just because none of our adaptations is able to proceed in this time.  For the very large running times, the Lin-Kernighan adaptations are outperformed by the state-of-the-art memetic algorithm which usually solves the problem to optimality.

To implement the most powerful adaptation `Exact', a new approach was proposed.  Note that the same approach can be applied to many other TSP local searches.  Comparing to the previous results in the literature, the time complexity of exploration of the corresponding neighborhood is significantly reduced which makes this adaptation practical.  Though it was often outperformed by either faster adaptations or the memetic algorithm in our experiments, it is clearly the best heuristic for small instances (up to 20 clusters in our experiments) and it is also naturally suitable for the asymmetric GTSP\@.

Further research on adaptation of the Lin-Kernighan heuristic for other combinatorial optimization problems may be of interest.  Our future plans also include a thorough study of different GTSP neighborhoods and their combinations.


\bibliographystyle{model2-names}
\bibliography{GtspLinKernighan}{}

\begin{table}[ht] \centering
\footnotesize
\caption{Details of experiment results for the small (10 to 29 clusters) instances.}
\label{tab:detailed_small}
\begin{tabular}{l @{\quad} r @{\quad} c *{5}{@{\hspace{1.5em}} r} @{\quad} c *{5}{@{\hspace{1.5em}} r}}
\toprule
&&& \multicolumn{5}{c}{Solution error, \%} && \multicolumn{5}{c}{Running time, ms} \\
\cmidrule(r){4-8} 
\cmidrule(r){10-14}
Instance&Best&&\optshort{2}{B}{co}&\LKc{5}{2}{co}&\LKs{5}{2}{co}&\LKe{4}{2}{}&\MA&&\optshort{2}{B}{co}&\LKc{5}{2}{co}&\LKs{5}{2}{co}&\LKe{4}{2}{}&\MA\\
\cmidrule(){1-14}

10att48&5394&&6.3&0.0&0.0&0.0&0.0&&0.24&0.25&0.28&2.53&18.72\\
10gr48&1834&&4.9&0.0&0.0&0.0&0.0&&0.01&0.03&0.06&1.45&12.48\\
10hk48&6386&&0.0&0.3&0.0&0.0&0.0&&0.01&0.08&0.17&1.36&18.72\\
11eil51&174&&4.0&0.6&0.0&0.0&0.0&&0.01&0.04&0.13&1.23&17.16\\
11berlin52&4040&&0.0&0.0&0.4&0.0&0.0&&0.01&0.06&0.12&1.17&12.48\\
12brazil58&15332&&2.1&0.0&0.0&0.0&0.0&&0.01&0.06&0.11&1.67&12.48\\
14st70&316&&6.3&0.0&0.3&0.0&0.0&&0.02&0.07&0.16&3.18&21.84\\
16eil76&209&&4.8&0.0&0.0&0.0&0.0&&0.01&0.06&0.23&4.23&21.84\\
16pr76&64925&&1.6&1.4&0.0&0.0&0.0&&0.02&0.11&0.27&4.10&26.52\\
20gr96&29440&&2.4&1.0&0.0&0.0&0.0&&0.03&0.22&0.42&9.09&28.08\\
20rat99&497&&7.8&0.2&0.0&0.0&0.0&&0.03&0.24&0.64&15.05&37.44\\
20kroa100&9711&&4.2&5.8&0.0&0.0&0.0&&0.03&0.17&0.46&14.59&31.20\\
20krob100&10328&&0.0&0.0&0.0&0.0&0.0&&0.01&0.10&0.36&15.64&28.08\\
20kroc100&9554&&10.1&0.1&0.0&0.0&0.0&&0.03&0.17&0.54&8.05&31.20\\
20krod100&9450&&1.5&0.0&2.0&0.0&0.0&&0.05&0.17&0.44&10.12&39.00\\
20kroe100&9523&&1.3&4.4&0.0&0.0&0.0&&0.03&0.15&0.37&8.33&31.20\\
20rd100&3650&&7.1&0.1&0.9&0.0&0.0&&0.03&0.15&0.53&18.02&34.32\\
21eil101&249&&4.4&0.4&0.8&0.4&0.0&&0.02&0.16&0.30&7.24&43.68\\
21lin105&8213&&0.1&0.0&0.0&0.0&0.0&&0.02&0.14&0.36&5.50&32.76\\
22pr107&27898&&4.4&0.0&0.0&0.0&0.0&&0.01&0.23&0.42&24.48&31.20\\
24gr120&2769&&20.5&2.8&2.6&0.0&0.0&&0.03&0.24&0.77&10.77&43.68\\
25pr124&36605&&4.5&0.0&0.5&0.0&0.0&&0.05&0.39&0.81&14.76&46.80\\
26bier127&72418&&6.9&8.6&0.0&0.0&0.0&&0.08&0.36&0.69&12.45&54.60\\
26ch130&2828&&12.1&0.0&0.0&0.0&0.0&&0.09&0.24&0.71&18.14&48.36\\
28pr136&42570&&9.7&0.8&0.0&0.0&0.0&&0.04&0.49&0.77&14.24&49.92\\
28gr137&36417&&1.9&1.4&1.3&0.1&0.0&&0.04&0.27&0.97&62.66&51.48\\
29pr144&45886&&4.0&0.0&0.0&0.0&0.0&&0.03&0.36&0.58&15.31&40.56\\
\cmidrule(){1-14}

Average&&&4.9&1.0&0.3&0.0&0.0&&0.04&0.19&0.43&11.31&32.07\\
\bottomrule

\end{tabular}
\end{table}

\begin{table}[ht] \centering
\footnotesize
\caption{Detailed experiment results for the moderate and large instances ($m \ge 30$).  The reported values are relative solution errors, \%.}
\label{tab:quality}
\begin{tabular}{l@{\hspace{1.5em}}r@{\hspace{1.5em}}c@{\hspace{1.5em}}r@{\hspace{1.5em}}r@{\hspace{1.5em}}r@{\hspace{1.5em}}r@{\hspace{1.5em}}r@{\hspace{1.5em}}r@{\hspace{1.5em}}r@{\hspace{1.5em}}r@{\hspace{1.5em}}r}
\toprule

Instance&Best&&\optshort{2}{B}{co}&\LKb{2}{2}{co}&\LKc{5}{2}{co}&\LKs{5}{2}{co}&\LKs{5}{3}{co}&\LKs{5}{4}{co}&\LKs{4}{4}{co}&\LKe{4}{2}{}&\MA\\
\cmidrule(){1-12}

30ch150&2750&&6.5&7.1&1.7&1.1&0.0&0.0&0.3&1.1&0.0\\
30kroa150&11018&&16.2&8.2&0.1&1.6&0.0&0.0&0.0&0.0&0.0\\
30krob150&12196&&5.4&5.4&0.0&1.0&0.0&0.5&0.0&0.0&0.0\\
31pr152&51576&&4.1&3.3&3.9&1.9&0.0&0.0&0.0&1.2&0.0\\
32u159&22664&&24.9&10.2&0.8&0.4&0.0&0.0&0.0&1.1&0.0\\
35si175&5564&&2.6&3.8&0.0&0.0&0.0&0.0&0.0&0.0&0.0\\
36brg180&4420&&314.5&314.5&0.5&78.3&0.0&0.0&0.0&0.0&0.0\\
39rat195&854&&7.6&12.5&1.4&2.0&0.2&1.3&0.1&0.0&0.0\\
40d198&10557&&1.3&3.5&1.3&0.3&0.0&0.5&0.0&0.2&0.0\\
40kroa200&13406&&8.3&4.8&0.6&0.4&0.4&0.4&0.4&0.0&0.0\\
40krob200&13111&&14.6&14.0&0.1&2.7&0.2&0.1&0.0&0.0&0.0\\
41gr202&23301&&10.5&7.1&3.1&4.3&2.5&1.9&0.0&0.0&0.0\\
45ts225&68340&&7.2&6.8&0.1&0.3&0.3&0.1&0.3&0.1&0.0\\
45tsp225&1612&&12.3&6.6&0.6&1.0&1.9&0.3&0.3&0.0&0.0\\
46pr226&64007&&14.2&1.1&1.1&1.1&0.0&0.0&0.0&0.0&0.0\\
46gr229&71972&&7.6&8.1&1.2&1.0&0.0&0.0&0.0&0.9&0.0\\
53gil262&1013&&20.6&11.1&3.6&0.7&0.8&0.7&0.0&0.2&0.0\\
53pr264&29549&&9.9&0.7&0.8&0.8&1.0&0.4&0.2&0.5&0.0\\
56a280&1079&&5.9&3.3&2.3&0.8&0.3&0.3&0.0&0.6&0.0\\
60pr299&22615&&8.0&4.0&3.9&1.0&0.2&0.0&0.0&0.1&0.0\\
64lin318&20765&&10.1&8.3&3.7&2.5&0.9&0.0&0.0&2.6&0.0\\
80rd400&6361&&11.4&7.9&2.3&1.3&2.8&1.1&2.0&0.7&0.0\\
84fl417&9651&&0.5&1.5&1.8&1.6&0.5&0.0&0.1&0.0&0.0\\
87gr431&101946&&5.1&5.2&2.6&3.2&3.6&2.5&1.1&0.0&0.0\\
88pr439&60099&&9.7&5.9&1.8&1.2&1.4&1.3&0.0&1.1&0.0\\
89pcb442&21657&&7.8&5.5&2.9&0.1&1.0&0.0&1.7&2.1&0.0\\
99d493&20023&&8.3&5.8&2.1&3.3&2.4&0.7&1.4&2.3&0.0\\
107ali535&128639&&15.9&5.0&3.1&2.6&0.5&0.0&0.4&0.5&0.0\\
107att532&13464&&11.3&5.6&0.8&1.5&0.5&0.8&0.1&0.1&0.0\\
107si535&13502&&2.4&1.3&0.3&0.1&0.1&0.0&0.0&0.3&0.0\\
113pa561&1038&&10.7&6.3&1.4&2.9&1.7&1.6&1.6&0.6&0.0\\
115u574&16689&&10.4&9.5&5.7&5.1&0.2&1.1&1.0&1.6&0.0\\
115rat575&2388&&13.4&11.5&4.4&4.2&3.5&3.2&3.2&1.3&0.2\\
131p654&27428&&2.0&1.4&0.3&2.5&0.2&0.0&0.0&0.2&0.0\\
132d657&22498&&10.6&9.5&4.6&3.9&1.7&1.6&0.5&1.9&0.1\\
134gr666&163028&&10.7&5.7&2.2&2.4&1.9&2.5&2.0&1.0&0.2\\
145u724&17272&&12.5&13.1&4.6&2.3&1.3&2.9&0.3&1.3&0.0\\
157rat783&3262&&19.7&12.9&4.7&2.9&3.5&0.3&1.6&1.3&0.1\\
200dsj1000&9187884&&14.8&8.9&4.3&4.4&0.8&1.5&1.9&2.6&0.1\\
201pr1002&114311&&16.3&8.4&3.6&0.2&0.2&1.5&0.8&0.1&0.2\\
207si1032&22306&&5.2&4.1&1.7&1.2&0.9&0.1&0.1&0.9&0.0\\
212u1060&106007&&13.7&9.0&3.6&2.3&1.8&1.7&2.1&0.7&0.2\\
217vm1084&130704&&12.4&8.2&3.1&3.0&2.2&2.0&2.1&1.8&0.3\\
\cmidrule(){1-12}

Average&&&17.1&13.9&2.2&3.6&1.0&0.8&0.6&0.7&0.0\\
Light avg.&&&11.4&7.6&2.5&2.0&1.1&0.9&0.7&0.8&0.0\\
Heavy avg.&&&42.3&41.3&0.8&10.6&0.3&0.1&0.1&0.3&0.0\\
\bottomrule

\end{tabular}
\end{table}
\begin{table}[ht] \centering
\footnotesize
\caption{Detailed experiment results for the moderate and large instances ($m \ge 30$).  The reported values are running times, ms.}
\label{tab:time}
\begin{tabular}{l@{\hspace{1em}}c@{\hspace{1em}}r@{\hspace{1em}}r@{\hspace{1em}}r@{\hspace{1em}}r@{\hspace{1em}}r@{\hspace{1em}}r@{\hspace{1em}}r@{\hspace{1em}}r@{\hspace{1em}}r}
\toprule

Instance&&\optshort{2}{B}{co}&\LKb{2}{2}{co}&\LKc{5}{2}{co}&\LKs{5}{2}{co}&\LKs{5}{3}{co}&\LKs{5}{4}{co}&\LKs{4}{4}{co}&\LKe{4}{2}{}&\MA\\
\cmidrule(){1-11}

30ch150&&0.1&0.1&0.5&1.4&2.8&2.7&8.0&46.7&56.2\\
30kroa150&&0.0&0.1&0.4&1.0&1.8&4.0&4.2&32.3&57.7\\
30krob150&&0.0&0.1&0.4&1.2&1.5&2.5&7.0&50.6&65.5\\
31pr152&&0.0&0.2&0.4&1.4&4.5&25.3&33.4&38.8&39.0\\
32u159&&0.1&0.1&0.3&0.9&2.7&4.2&25.7&31.9&62.4\\
35si175&&0.1&0.2&1.8&3.6&10.0&23.5&358.8&232.5&64.0\\
36brg180&&0.0&0.2&0.4&0.4&1.2&2.3&279.3&46.4&53.0\\
39rat195&&0.1&0.1&0.7&1.2&3.1&7.3&13.7&64.9&138.8\\
40d198&&0.2&0.6&2.0&3.7&21.9&134.2&310.4&98.7&126.4\\
40kroa200&&0.1&0.1&0.7&1.6&3.2&4.1&11.7&60.6&123.2\\
40krob200&&0.1&0.2&0.5&1.4&2.4&4.2&16.2&56.3&157.6\\
41gr202&&0.1&0.3&0.8&1.9&7.8&11.0&81.2&86.1&198.1\\
45ts225&&0.1&0.3&0.7&3.0&8.0&10.0&19.9&273.0&191.9\\
45tsp225&&0.1&0.2&0.8&2.3&3.1&7.6&15.5&112.3&156.0\\
46pr226&&0.1&0.4&1.0&1.9&4.5&12.7&21.7&44.1&95.2\\
46gr229&&0.1&0.2&1.0&3.2&3.5&8.8&13.9&145.1&224.6\\
53gil262&&0.2&0.3&1.9&3.8&7.9&9.2&21.3&107.8&290.2\\
53pr264&&0.2&1.0&5.7&6.5&66.2&282.4&505.4&230.9&204.4\\
56a280&&0.2&0.3&1.1&2.2&11.2&9.3&43.9&148.2&291.7\\
60pr299&&0.1&0.2&1.5&3.8&8.7&12.6&31.4&146.7&347.9\\
64lin318&&0.2&0.3&2.0&4.2&17.3&48.6&81.4&223.1&404.0\\
80rd400&&0.3&0.7&3.8&5.6&18.2&36.7&74.4&305.8&872.0\\
84fl417&&0.3&2.3&5.9&9.7&59.0&174.8&315.1&645.8&583.4\\
87gr431&&0.4&0.8&4.4&9.3&19.8&59.6&107.9&485.2&1673.9\\
88pr439&&0.3&0.8&3.0&11.6&24.4&54.3&109.3&764.4&1146.6\\
89pcb442&&0.5&0.8&4.1&9.5&23.1&42.9&88.8&656.8&1530.4\\
99d493&&0.7&2.0&7.5&13.1&148.3&2666.1&1616.2&591.2&3675.4\\
107ali535&&1.0&2.3&7.1&13.4&29.9&52.2&170.2&795.6&3558.4\\
107att532&&0.6&1.9&8.0&17.1&33.1&71.5&312.1&932.9&2942.2\\
107si535&&0.5&5.5&32.9&46.7&337.0&1921.9&12725.0&3503.8&1449.2\\
113pa561&&0.7&1.3&5.4&11.6&28.6&51.0&104.2&695.8&2931.3\\
115u574&&0.7&1.9&6.8&10.7&53.3&63.9&156.1&956.3&3017.1\\
115rat575&&0.5&1.3&6.4&17.9&41.0&92.9&128.0&697.3&2867.3\\
131p654&&1.2&9.4&40.9&27.3&213.9&1074.8&2964.0&3293.2&2137.2\\
132d657&&0.9&2.3&13.6&22.0&109.4&1009.3&2322.9&794.0&4711.2\\
134gr666&&1.0&2.3&8.7&28.1&51.5&135.9&374.4&1425.8&10698.6\\
145u724&&1.0&2.7&13.4&32.6&62.8&105.8&242.0&1326.0&7952.9\\
157rat783&&1.5&2.2&17.7&30.8&73.7&131.3&248.3&2165.3&9459.9\\
200dsj1000&&3.5&10.3&80.7&104.5&592.8&5199.5&8032.5&9361.6&22704.4\\
201pr1002&&2.3&6.2&39.1&57.0&156.4&290.6&539.8&2719.1&21443.9\\
207si1032&&3.5&37.4&839.3&875.2&7063.7&195644.0&306944.8&112926.4&17840.3\\
212u1060&&3.7&7.1&36.4&80.2&195.5&307.5&1040.5&2990.5&31201.8\\
217vm1084&&2.5&6.6&51.4&78.5&204.8&496.1&978.1&4687.8&27587.2\\
\cmidrule(){1-11}

Average&&0.7&2.6&29.3&36.3&226.4&4890.9&7941.8&3604.6&4310.1\\
Light avg.&&0.7&1.6&9.5&16.8&56.0&315.7&488.5&972.0&4653.6\\
Heavy avg.&&0.8&7.1&116.1&121.6&971.6&24907.3&40550.4&15122.2&2807.2\\
\bottomrule

\end{tabular}
\end{table}

\end{document}